\documentclass[%
 aip,
 amsmath,amssymb,
 reprint,%
]{revtex4-1}

\usepackage{graphicx}
\usepackage{dcolumn}
\usepackage{bm}

\usepackage[utf8]{inputenc}
\usepackage[T1]{fontenc}
\usepackage{mathptmx}
\usepackage{etoolbox}

\makeatletter
\def\@email#1#2{%
 \endgroup
 \patchcmd{\titleblock@produce}
  {\frontmatter@RRAPformat}
  {\frontmatter@RRAPformat{\produce@RRAP{*#1\href{mailto:#2}{#2}}}\frontmatter@RRAPformat}
  {}{}
}%
\makeatother
\begin{document}

\preprint{AIP/123-QED}

\title{Charge-Dependent Coordination and Redox Energetics of Actinide Fluoride Complexes in Molten Salts}
\author{G. Bogolyubov}
 \email{gbogolyubov@gmail.com.}
\affiliation{Independent}%

\date{\today}

\begin{abstract}
Redox control in molten fluoride salt reactors determines actinide solubility and corrosion of structural alloys, but the role of local fluoride coordination in this control remains poorly understood. We use density functional theory with explicit LiF microsolvation to study the coordination chemistry, intrinsic redox energetics, and fluoride-exchange kinetics of Th(IV), U(IV), and U(III) fluoride complexes. The coordination response to solvation primarily follows the formal charge of the complex rather than the actinide at its center: we find that cationic and neutral complexes gain fluoride ligands, most anionic complexes keep their coordination number, and $4-$ complexes lose fluoride. Coordination number then strongly controls the intrinsic U(IV)/U(III) reduction energy. From CN~4 to CN~7, this energy rises by up to $11$~eV and changes sign; a continuum-solvation estimate opposes this trend and therefore establishes the intrinsic value as an upper bound. Fluoride exchange barriers are low, so redox partners of different charge do not keep the same coordination numbers in solvated clusters. Finally, above a threshold between CN~5 and CN~6, single-reference DFT converges to a ligand-reduced U(III) state inconsistent with formal $5f^3$ occupation, however the physical nature of this is still unresolved. Therefore, local coordination is a primary variable in the prediction of redox behavior in molten actinide-fluoride salts.
\end{abstract}

\maketitle

\section{Introduction \& Background}\label{sec:Introduction}
A molten salt reactor (MSR) dissolves actinide fuel directly into a molten fluoride carrier salt; the same liquid acts as both fuel and coolant \cite{WANG2024163}. This is what makes continuous removal of fission products and online reprocessing possible in MSRs: bred $^{233}$U and accumulated fission products can be handled simultaneously, allowing the reactor to achieve higher burnup and reducing the need for fresh uranium enrichment \cite{leblanc2010molten, serp2014molten, BENES2012359}. In the thorium-uranium (Th-U) fuel cycle, fertile $^{232}$Th captures a neutron, following the breeding sequence
\begin{equation}
{}^{232}\mathrm{Th}(n,\gamma){}^{233}\mathrm{Th} \xrightarrow{\beta^-}{}^{233}\mathrm{Pa} \xrightarrow{\beta^-}{}^{233}\mathrm{U},
\end{equation}
but the resulting fuel exists in the melt as fluoride-coordinated ionic complexes rather than isolated actinide ions. Depending on salt composition, temperature, and the local fluoride environment, thorium and uranium can form species such as $[\mathrm{ThF}_8]^{4-}$, $[\mathrm{UF}_8]^{4-}$, and $[\mathrm{UF}_7]^{3-}$. The chemically important control reaction is the uranium redox buffer,
\begin{equation}
\mathrm{U(IV)} + e^- \rightleftharpoons \mathrm{U(III)},
\end{equation}
which determines whether the salt is sufficiently oxidizing to dissolve structural metals as fluorides or sufficiently reducing to precipitate actinide metal. Reactor operation therefore depends both on nuclear transmutation and on the local coordination and redox chemistry of dissolved actinides. In practice, however, this Th-U fuel cycle presents several chemical and engineering challenges in an MSR. These include corrosion of structural materials, control of actinide solubility, and management of the salt redox state \cite{zhang2018redox, gibilaro2015way, baes1974chemistry}. The redox state of the fuel salt is an important control variable because it influences which actinide oxidation states and fluoride-coordinated species are thermodynamically stable. Controlling the redox state can therefore tune the distribution of species such as $[\mathrm{ThF}_8]^{4-}$, $[\mathrm{UF}_8]^{4-}$, and $[\mathrm{UF}_7]^{3-}$ \cite{zhang2018redox}. This distribution must remain within a narrow operating window: if the salt is too oxidizing, it oxidizes structural alloys and stabilizes their dissolved fluoride species, worsening corrosion. On the other hand, if the salt is too reducing, actinide fluoride species can be reduced to metallic uranium or thorium, which can precipitate and clog fuel channels, degrading heat transfer \cite{zhang2018redox, gibilaro2015way, baes1974chemistry}. The $\mathrm{U}^{4+}/\mathrm{U}^{3+}$ ratio is commonly used as a redox buffer to maintain this window \cite{zhang2018redox,baes1974chemistry}, but it depends on how the actinides are coordinated by fluoride in the melt. Therefore, fuel solubility, alloy corrosion, and redox control are all linked through local actinide coordination; quantifying that connection is the goal of this study.

What is currently known about actinide-fluoride coordination in molten fluoride salts is incomplete. Here, the coordination number (CN) is the number of fluoride ligands in the first coordination shell around an actinide center. In general, fluoride complexation can be represented as
\begin{equation}
\mathrm{An}^{z+} + n\mathrm{F}^{-} \rightleftharpoons [\mathrm{AnF}_{n}]^{(z-n)+},
\end{equation}
where $\mathrm{An}$ is some actinide, $z$ is its oxidation state, and $n$ is the coordination number. High-temperature structural measurements by Smith and Bessada have determined average actinide-fluoride coordination numbers in molten fluoride salts, but they have not systematically explained why Th(IV), U(IV), and U(III) prefer particular coordination states or how their electronic structures determine those preferences \cite{smith2019exafs, smith2022structure, bessada2017high, bessada2020investigation}. Our best idea of what goes on in the salt comes from experimental measurements of thorium. Extended X-Ray Absorption Fine Structure (EXAFS) measurements of molten $\mathrm{AF}$-$\mathrm{ThF}_{4}$ mixtures ($\mathrm{AF}$ represents an alkali fluoride such as LiF, NaF, or KF) find thorium-fluoride coordination numbers from 6 to 10; the dominant thorium species depends on the alkali cation \cite{bessada2017high, smith2019exafs, bessada2020investigation}. Smith \textit{et al.} measured the mean number of fluoride ligands coordinated to dissolved Th(IV) and found that it decreases from 8.1 at 1273 K to 7.6 at 1733 K. As the melt's temperature increases, the distribution broadens from predominantly 8- and 9-coordinate thorium species to a mixture of 6- through 9-coordinate species, consistent with a more labile fluoride coordination shell \cite{smith2022structure, dai2015molecular}. U(IV) behaves similarly in comparable fluoride melts, with average U-F coordination numbers of 7.4 to 7.9. At equal coordination number, its U-F bonds are 0.03 to 0.05 \AA{} shorter than the corresponding Th-F bonds, an indication of actinide contraction \cite{bessada2020investigation, smith2022structure, ocadiz2021new}. U(III), the reduced member of the $\mathrm{U}^{4+}/\mathrm{U}^{3+}$ redox buffer, matters most directly for redox control because its population relative to U(IV) determines the oxidation state of the fuel salt. At fixed coordination, this reduction can be written as
\begin{equation}
[\mathrm{UF}_{n}]^{(4-n)+} + e^- \rightleftharpoons [\mathrm{UF}_{n}]^{(3-n)+}.
\end{equation}
In the real melt, however, the coordination number can change as the complex exchanges fluoride with its surroundings. Consequently, there is very little direct experimental information on the coordination or redox chemistry of U(III) complexes in molten fluorides.

For the most part, these structures have been reproduced by calculation. Unfortunately, the calculations are not able to fully characterize them. Molecular dynamics with polarizable ion models (MD-PIM), for example, recovers [ThF$_8$]$^{4-}$ and [ThF$_7$]$^{3-}$ as the dominant species in AF-ThF$_4$ melts, and DFT has found octahedral [ThF$_6$]$^{2-}$ and capped trigonal-prismatic [ThF$_7$]$^{3-}$ to be stable geometries \cite{liu2014theoretical, dai2015molecular, wang2021structures}. The trouble is that classical MD can't see electronic-structure effects or tell one oxidation state from another, and the DFT studies so far have looked at complexes one at a time instead of comparing across oxidation states. The redox side is in worse shape. Although reactor-scale tools such as SaltProc follow the bulk redox potential using electrochemical measurements and empirical correlations \cite{rykhlevskii2019modeling, yardas2023implementation}, they can't actually predict how $E^\circ$ will shift when the salt composition changes, when the temperature swings during startup or shutdown, or when fission products build up. There is also no mechanistic picture of fluoride-exchange kinetics, even though every redox step runs through the making and breaking of U-F and Th-F bonds. Those exchange barriers decide how quickly the salt reaches equilibrium and how cleanly actinides can be separated during reprocessing. Because of these gaps in chemistry, reactors are designed with larger redox-control margins than they may actually need, which means giving up some fuel utilization and reprocessing efficiency. However, there are solid experimental measurements to work from. Straka \textit{et al.} measured the U$^{4+}$/U$^{3+}$ couple in LiF-BeF$_2$ at around $-0.60$ to $-0.80$ V (vs. Ni$^{2+}$/Ni at 540$^{\circ}$C) \cite{straka2010electrochemistry}, and Jiang \textit{et al.} found reversible, diffusion-controlled U(IV)/U(III) behavior in FLiNaK \cite{jiang2017electrochemical}. The only systematic thorium measurement comes from Clayton \textit{et al.}, who put the Th$^{4+}$/Th couple near $-2.13$ V in LiF-NaF-KF \cite{clayton1974electrochemical}. Moreover, EXAFS has given us some idea of structure: U$^{4+}$ sits at the higher coordination numbers (CN $\approx$ 7-8, U-F $\approx$ 2.27-2.28 \AA), and U$^{3+}$ at the lower ones (CN $\approx$ 6) \cite{smith2022structure}. But as Hege \textit{et al.} point out in their 2023 review, we know far less about redox speciation in molten salts than in aqueous systems \cite{hege2023fundamental}. Nobody has yet computed the reduction energetics of these couples from electronic structure, connected coordination to redox thermodynamics mechanistically, or worked out the barriers to fluoride exchange, for either actinide.

This is the gap this study sets out to fill, starting from the electronic structure. Using relativistic density functional theory, we build a systematic series of Th(IV), U(IV), and U(III) fluoride complexes with coordination numbers from 3 to 8 (4 to 7 for U(III)), treated both bare and inside explicit salt environments built up from as many as eight LiF ion pairs. We compute free energies, reduction energetics, and fluoride-exchange transition states across 298, 973, and 1273~K. We set out to answer the following questions:
\begin{enumerate}
    \item How does U(III) coordination chemistry differ from that of Th(IV) and U(IV) in fluoride carrier salts, and what does that mean for redox processes in MSRs?
    \item What coordination numbers are thermodynamically preferred for each actinide, and how do first- and second-shell LiF solvation environments shift those preferences away from the bare complexes?
    \item Which electronic-structure factors set the relative stability of geometries across oxidation states, how do they carry through into redox energetics and exchange kinetics, and where does a single-determinant description of them break down?
\end{enumerate}
This study is structured as follows: section \ref{sec:Compdetails} covers the computational methodology, Section \ref{sec:Results} presents the results and the electronic-structure analysis, and Section \ref{sec:Discussion} goes over the implications for MSR fuel chemistry and future computational actinide work.

\section{Computational Methods}
\label{sec:Compdetails}
 
Every electronic-structure property reported here was calculated using the PBE0 hybrid functional \cite{adamo1999toward} with the scalar-relativistic ZORA Hamiltonian except for the multireference analysis of Section~\ref{sec:multiref}. This was so bare and solvated species at all three oxidation states can be compared without energy conversion. PBE0 was chosen as a strong candidate functional due to its 25\% exact Hartree-Fock exchange (this reduces the self-interaction error that is prominent in the open $5f$ shell of the actinides \cite{wen2013density}). Aebersold and Wilson \cite{aebersold2021considering} have shown TPSS can outperform hybrids for certain actinide properties, but PBE0 is the better validated choice for An-F coordination and in any case held up consistently across Th(IV), U(IV), and U(III) without the state-by-state retuning another functional would have needed. The zeroth-order regular approximation (ZORA) \cite{VANLENTHE2000107} was used to treat scalar relativistic effects, and every open-shell species was described in the spin-unrestricted Kohn-Sham formalism. Thorium and uranium used the all-electron, relativistically contracted SARC-ZORA basis set \cite{pantazis2011all}. These are built for scalar-relativistic ZORA work on the heavy elements and run through $f$ angular momentum with no $g$ polarization functions, as is appropriate for density functional rather than correlated wavefunction calculations \cite{pantazis2011all}. Fluorine and lithium were given the larger def2-QZVP basis \cite{weigend2005balanced}, since the properties studied depend mainly on the flexibility of the fluoride valence region (which a quadruple-$\zeta$ ligand basis describes well) while the actinide $5f$ and $6d$ manifolds are already well described at the SARC level. The RIJCOSX approximation \cite{NEESE200998} and the def2/J auxiliary basis \cite{weigend2006accurate} were used for the Coulomb and exchange terms on the program's default integration grid (\texttt{DefGrid2}, OptM3 radial grid). ORCA version 6.1.0 was used at all times \cite{neese2012orca,neese2025orca6}.
 
The bare $[\text{AnF}_n]^{m\pm}$ complexes were optimized and fully characterized at this PBE0/ZORA/SARC-ZORA level, including their analytic harmonic frequencies. For the microsolvated clusters their size makes a full PBE0/ZORA optimization computationally expensive and led to several convergence failures, especially for the open-shell clusters. Therefore, their geometries and frequencies were computed in Gaussian~16 at UB3LYP \cite{becke1993new,lee1988development} with the Stuttgart-Dresden (SDD) relativistic effective-core-potential basis on the actinide and the associated all-electron basis on F and Li, under tight SCF convergence and an ultrafine grid; this converged reliably and much faster than the equivalent all-electron ORCA optimizations. Total and relative energies, Mulliken and L\"owdin charges, reduced $s/p/d/f$ orbital populations, and Mayer bond orders were then computed as PBE0/ZORA/SARC-ZORA single points in ORCA on those geometries with thermal and entropic corrections taken from the UB3LYP/SDD frequencies within a quasi-rigid-rotor/harmonic-oscillator (quasi-RRHO) treatment of the soft, low-frequency modes. That way the quantities interpreted in Sections~\ref{sec:ucomp}-\ref{sec:redox} are reported uniformly at the PBE0/ZORA level, on UB3LYP/SDD geometries throughout the solvated series. However, mean An-F distances found at UB3LYP/SDD exceed those from an equivalent PBE0/ZORA optimization by $0.01$-$0.02$~\AA{} for the closed-shell Th(IV) species and by up to $0.14$~\AA{} for U(III), so the two levels are not interchangeable; the exchange barriers of Section~\ref{sec:kinetics} are the one exception to the PBE0/ZORA property level. The geometries and frequencies define a surface where the transition state is found, so naturally they must be UB3LYP/SDD energies on UB3LYP/SDD surfaces, and they are compared with each other instead of against the PBE0/ZORA energetics.
 
All ORCA calculations were converged to the tight criterion (\texttt{TightSCF}, roughly $10^{-8}$~a.u.\ in the energy). The closed-shell Th(IV) complexes converged nicely under standard DIIS extrapolation, but the open-shell uranium species ($5f^2$ triplets for U(IV), $5f^3$ quartets for U(III)) were quite stubborn, which is not unexpected with several near-degenerate $5f$ configurations crowding the Fermi level. Most of them converged under DIIS with slow-convergence damping (\texttt{SlowConv}), a level shift of $0.3$-$0.5$~$E_\mathrm{h}$, and an extended iteration limit; a few really problematic cases needed the automatic trust-region augmented Hessian procedure (\texttt{AutoTRAH}), and a handful ran to several thousand iterations before converging. One species, $[\text{UF}_7]^{4-}\!\cdot\!8$LiF, could not be converged under the trust-region procedure at all, the orbital rotations pushing the energy through positive values and generating an unphysical orbital spectrum; it converged under plain DIIS with a $1.0$~$E_\mathrm{h}$ level shift. Spin multiplicities were fixed from the formal $5f$ occupation (singlet for Th(IV), triplet for U(IV), quartet for U(III)) but the multiplicity alone does not necessarily guarantee the intended state, especially when dealing with open-shell uranium centers. We therefore verified every open-shell species from the uranium Mulliken spin population (U(IV) $\approx 2.0$, U(III) $\approx 3.0$) rather than from $\langle S^2\rangle$, which can and did give us a clean quartet-like value even when the metal has actually collapsed to an $f^2$ state. In fact, the metal spin population is the appropriate check here, not the global $\alpha$-$\beta$ electron count; in the spin-unrestricted formalism at fixed multiplicity the latter is constrained to $2S$ by construction. The deviations of up to $0.8$~e that appear in the printed values for the most highly charged anions track the corresponding error in the integrated total electron count almost exactly, and so we interpret them as artifacts of numerical integration rather than spin delocalization onto the ligands. Species that converged to a state inconsistent with their formal $5f$ occupation were excluded from the coordination and thermodynamic analysis; the collapsed solvated U(III) clusters are instead characterized directly in Section~\ref{sec:multiref}. The spin populations and $\langle S^2\rangle$ values are reported in the Supporting Information.
 
Symmetry was not imposed on equilibrium structures, so each complex relaxed to the nearest local minimum, and the structures were converged to ORCA's tight geometry thresholds (bare) or Gaussian's default thresholds (clusters). The more weakly bound, low-frequency species produced very soft modes, so we accepted structures with $|\omega_\mathrm{min}| < 30$~cm$^{-1}$ as minima, interpreting the smaller frequencies as numerical noise rather than actual imaginary modes. Two solvated species fall into this category. [UF$_6]^{3-}\!\cdot8\text{LiF}$ retains two residual modes, at $-7.1$ and $-4.9~cm^{-1}$, and [UF$_7]^{4-}\!\cdot8\text{LiF}$ three, at $-10.4$, $-6.0$, and $-3.6~cm^{-1}$; all five are well below the acceptance threshold. Every other structure in the solvated series is a clean minimum and the individual frequencies are noted in the Supporting Information. One bare species, [UF$_3$]$^+$, was not able to be characterized at a single-reference level: its frequency calculation failed to converge at 973~K and returned a second-order saddle at 1273~K. Its 298~K value is noted, but the unconverged 973 and 1273~K free energies are not (Table~\ref{tab:relG}, Table~S9), and [UF$_3$]$^+$ is dropped from the coordination-preference comparison at those two temperatures.  

There were three changes to the standard method for microsolvated optimizations when the standard failed. Firstly, there were several clusters that could not be optimized in redundant internal coordinates; when an An$\cdots$F-Li chain approaches linearity, the internal-coordinate construction becomes singular, and so these species were reoptimized in Cartesian coordinates from the last geometry reached. Secondly, the solvated U(III) clusters showed an SCF that alternated between two solutions differing by roughly $0.2$~$E_\mathrm{h}$ along the optimization path, so that the gradient was evaluated on different electronic states at successive steps. Therefore, a level shift of $0.3$~$E_\mathrm{h}$ was applied to hold a single state, without which no geometry could be located for these species. Thirdly, three clusters converged only to Gaussian's loose thresholds. In these cases the RMS force falls within the default criterion and the maximum force within a factor of two of it, while the displacement criteria fail on the outer Li positions; this is usually a sign of a flat potential surface, implied by the low exchange barriers of Section~\ref{sec:kinetics}, and we regard tighter displacement convergence on the outer shell as a precision the surface does not have. The species affected by these are identified in the Supporting Information. Finally, for every species whose initial optimization terminated without relaxing, a control optimization was run from a starting geometry displaced by $0.03$~\AA{} in all Cartesian coordinates. These converged to the same minimum as the corresponding base optimization in all cases but one, agreeing in mean An-F distance to better than $0.005$~\AA{}; the exception is $[\text{UF}_7]^{4-}\!\cdot\!8$LiF, for which the displaced start located a distinct structure $1.5$~eV lower in energy. That lower-energy structure is the one reported.
 
Natural bond orbital analysis used NBO 7.0 \cite{nboglend7}, giving natural population analysis (NPA) charges, natural localized molecular orbitals, and Wiberg/Mayer bond indices through the \texttt{\$NBO NPA NLMO BNDIDX \$END} directive. Mulliken, L\"owdin, and reduced $s/p/d/f$ analysis, came from the standard ORCA output. The L\"owdin charges are unreliable for the diffuse SARC-ZORA actinide basis at high coordination, where the orthogonalization misassigns diffuse density to the metal and provides unphysical negative charges; we therefore report Mulliken charges for $[\text{UF}_7]^{3-}$ and $[\text{UF}_8]^{4-}$. These negative values are reproduced by every functional tested (Supporting Information), which confirms their origin in the basis and the orthogonalization scheme rather than in the density. Because the all-electron SARC-ZORA basis carries the filled $4f^{14}$ semicore explicitly, the reduced $f$ populations quoted below are total $4f+5f$ populations rather than $5f$ populations alone, and run near $14$ for thorium and near $16$-$17$ for uranium. Two complexes take cubic point groups in their optimized structures: [ThF$_4$] ($T_d$) and [ThF$_6$]$^{2-}$ ($O_h$). The analysis returns problematic populations on orbitals spanning degenerate irreducible representations. We applied a $0.001$~\AA{} random Cartesian displacement to a single halide to remove the symmetry-induced degeneracy while leaving the electronic structure essentially untouched ($\Delta E < 10^{-5}~E_\mathrm{h}$); the resulting populations were insensitive to the displacement magnitude across the range tested. Frontier-orbital energies and HOMO-LUMO gaps were taken from the converged Kohn-Sham orbitals. For the open-shell uranium species the gap is defined across both spin channels; HOMO the highest occupied and LUMO the lowest virtual orbital over the combined $\alpha$ and $\beta$ sets, not within a single channel. The uranium LUMO usually lies in the $\beta$ channel, and an $\alpha$-only gap would skip that lower-lying virtual and overstate the gap badly (for $[\text{UF}_7]^{4-}$, for example, by more than $4$~eV). We note too that the HOMO energies of the most highly anionic gas-phase species (the $2-$, $3-$, and $4-$ ions) run strongly positive, as one expects for highly charged anions bound only by the finite basis; this shifts the absolute orbital energies but does not affect the relative gap trends across coordination number.
 
To quantify how the molten-salt environment affects An-F coordination, each complex was treated at three levels of solvation: the bare [AnF$_n$]$^{m-}$ ion, a first shell [AnF$_n$(LiF)$_4$]$^{m-}$, and a combined first and second shell [AnF$_n$(LiF)$_8$]$^{m-}$. The solvating Li$^+$/F$^-$ pairs were placed around the optimized actinide-fluoride core following MD \cite{liu2014theoretical,DEWAN2013322} and EXAFS \cite{smith2019exafs} data for the melt. Each Li$^+$ sat at $\langle r(\mathrm{An\text{-}F})\rangle + 1.84$~\AA{} from the metal - the first term the mean An-F bond length of the bare complex, the second the most probable Li$^+$-F$^-$ separation in the melt - with its partner F$^-$ placed a further $2.0$~\AA{} out along the same radial line. The four first-shell pairs were arranged to keep stoichiometric coordination about both the actinide and the surrounding Li$^+$ centers, and the second shell added four more pairs at larger radius to capture longer-range electrostatic screening. All atoms relaxed freely with each cluster settling into the local minimum nearest to its starting arrangement. Basis-set superposition error was not corrected in the solvated energies. Every solvated quantity interpreted here is a difference taken at fixed oxidation state, fixed solvation level, and fixed basis (coordination-number preferences within a series, and coordination-number changes on solvation) and the counterpoise correction for a given actinide-fluoride core in a given LiF environment is nearly constant across such a comparison and cancels to first order. The absolute binding energy of a cluster relative to its separated fragments, for which the correction would be required, is not a quantity we report. The correction is in any case poorly defined for this series: several clusters draw a fluoride out of the LiF shell into the first coordination sphere during relaxation, so the core/shell fragment partition on which the Boys-Bernardi construction depends is not the same for all species, and no uniform correction could be applied across the set. Solvation here is explicit throughout - the melt is represented by discrete Li$^+$/F$^-$ counter-ions, and no dielectric continuum was applied at any stage. Relative enthalpies and free energies were computed at three temperatures: 298.15~K, for reference with standard thermodynamic tables; 973.0~K, representative of FLiBe/FLiNaK operation; and 1273.0~K, an upper bound approaching peak FLiBe coolant temperatures \cite{busby2019technical,wna_msr}. Thermal corrections came from the harmonic frequencies within the (quasi-)RRHO ideal-gas approximation, the $973$ and $1273$~K values obtained from the same Hessian as the $298$~K result. This introduces a systematic error in the absolute free energies, but the comparisons we make in our conclusions are between coordination numbers at fixed oxidation state, where that error largely cancels. Gibbs free energies followed as $\Delta G(T) = \Delta H(T) - T\Delta S(T)$, and the relative free energies of the coordination numbers were referenced to the most stable species at each temperature with stoichiometric balancing by free fluoride, the free-fluoride free energy taken from the Sackur-Tetrode equation (Supporting Information).
 
For each U(IV)/U(III) couple we characterize the one-electron half-reaction U(IV) $+\,e^- \rightarrow$ U(III) by its adiabatic energetics: with both oxidation states relaxed at a common coordination number and environment, the adiabatic reduction energy is $\Delta E_\mathrm{adiab} = E[\text{U(III)}_\mathrm{relaxed}] - E[\text{U(IV)}_\mathrm{relaxed}]$, and the corresponding free energy is found by adding the frequency-derived thermal and entropic corrections. These are the intrinsic reduction energetics of the couple; we report them as computed, without referencing to an absolute electrode such as the standard hydrogen electrode, so the values are reduction energies rather than potentials on the electrochemical scale. To actually place them on that scale we would need to include an absolute-SHE offset (approximately $4.3$~V) which would shift the whole series into the range of measured molten-salt potentials; because that offset is a single additive constant, it does not affect the coordination- and environment-dependent trends of this study. We therefore compare the electrochemical measurements of Straka \textit{et al.} \cite{straka2010electrochemistry}, Jiang \textit{et al.} \cite{jiang2017electrochemical}, and Clayton \textit{et al.} \cite{clayton1974electrochemical} with those trends rather than absolute volts. As a reproducibility check the reduction energies were re-derived from an independent set of single points on the same optimized geometries, agreeing with the tabulated values to $0.01$~eV at every coordination number recovered (Supporting Information). 

The inner-sphere reorganization energy of each couple was found from the standard four-point construction, evaluating each oxidation state at both its own and its counterpart's relaxed geometry \cite{marcus1985electron}. Writing $E_\mathrm{ox}$ and $E_\mathrm{red}$ for the oxidized (U(IV)) and reduced (U(III)) surfaces and $R_\mathrm{ox}$, $R_\mathrm{red}$ for their optimized geometries,
\begin{equation}
\begin{aligned}
\lambda_i = \frac{1}{2}\bigl[&
E_\mathrm{ox}(R_\mathrm{red}) - E_\mathrm{ox}(R_\mathrm{ox}) \\
&+ E_\mathrm{red}(R_\mathrm{ox}) - E_\mathrm{red}(R_\mathrm{red}) \bigr].
\end{aligned}
\label{eq:lambda}
\end{equation}
each bracketed term being the energy cost of holding one redox state at the other's geometry. The geometric reorganization of the first coordination shell is computed for the bare ions and with no continuum contribution, so no outer-sphere dielectric term enters. The two cross-geometry single points required for each couple were seeded (\texttt{MOREAD}) from a clean-quartet neighbor rather than from the native SCF guess, since the native high-coordination U(III) guess tends to fall into the wrong electronic state, and the spin state of each resulting point was then examined with the same population gate used above. The corresponding construction in the microsolvated clusters was attempted and abandoned, since the cross-geometry points there either failed to converge or converged to the collapsed state described below, so solvated reorganization energies are not reported. We report the bare values as an observation rather than a result, for the reason given in Section~\ref{sec:redox}.
 
Fluoride-exchange kinetics were characterized in Gaussian~16 at UB3LYP/SDD, matching the level used for the cluster geometries. For each exchange we ran relaxed scans along the An-F coordinate, followed by transition-state optimization and frequency verification; each transition state was confirmed by a single imaginary frequency whose eigenvector corresponds to the An-F exchange motion. Scan step counts and ranges are given in the Supporting Information. One transition state, for the [UF$_7$]$^{3-}\rightarrow$[UF$_8$]$^{4-}$ exchange, had an unusually large imaginary frequency ($-570$~cm$^{-1}$) but remained a genuine first-order saddle with a single imaginary mode along the exchange coordinate.
 
The solvated U(III) clusters keep their formal $f^3$ configuration at low coordination and lose it at high coordination, with the transition falling between CN~5 and CN~6. Reading the uranium Mulliken spin population from the converged PBE0/ZORA densities on the first-shell clusters, CN~4 and CN~5 are clean quartets ($3.10$ and $3.08$), CN~6 has collapsed to a ligand-reduced $f^2$ state ($2.34$), and CN~7 lies between the two ($2.60$). The second shell follows the same pattern where it is characterized; CN~5 remains a clean quartet ($3.08$) while CN~6 and CN~7 are fully collapsed ($2.08$ in both cases). Where the collapse occurs the third electron delocalizes onto the fluoride and LiF environment. The solvated reduction and reorganization energies of the affected species are accordingly withheld; the origin of the collapse is discussed in Section~\ref{sec:redox} and its multireference character in Section~\ref{sec:multiref}.
 
The collapsed clusters were examined further with a multireference treatment, carried out on the $+8$LiF UB3LYP/SDD geometries of the CN~5, CN~6, and CN~7 U(III) species. Because the uranium $5f$ manifold in these clusters is fragmented across the canonical orbital spectrum, no active space can be selected by orbital energy; we therefore constructed it from atomic-valence character using the atomic-valence active-space (AVAS) procedure \cite{sayfutyarova2017automated}, projecting the converged PBE0/ZORA orbitals onto the seven uranium valence $f$ shells with an ANO-RCC minimal reference basis and a projection threshold of $0.25$. At CN~5 and CN~6 this recovers a CAS($3,7$) quartet, three electrons in seven orbitals of predominantly U-$5f$ character. At CN~7 the same construction selects seven orbitals but only one electron, because six of the seven $5f$-character orbitals lie above the Fermi level in the converged density; a CAS($3,7$) cannot be built there, and we report that failure rather than forcing the selection. Lowering the projection threshold to $0.10$ at CN~7 admits ligand character and yields a CAS($5,9$) instead, which we report separately and do not compare directly against the CAS($3,7$) results. Configuration-interaction energies and wavefunctions were obtained in each fixed active space without orbital optimization, so that active-space natural-orbital occupations were not available as a second gate: the electron count is imposed by the AVAS construction, and the uranium spin population remains the gate on metal localization. Determinant weights were printed down to a threshold of $0.01$, and configurations are counted below at a weight above $0.05$. Orbital optimization was attempted in every active space that could be constructed, from the converged CASCI orbitals, under three settings: undamped, and with level shifts of $2.0$ and $3.0$~$E_\mathrm{h}$ paired with maximum orbital rotations of $0.03$ and $0.015$ respectively. All ran to a macroiteration cap of $200$ and none converged; the trajectories are described in Section~\ref{sec:multiref}.
 
The sensitivity of the reported properties to the choice of functional was assessed by single-point recalculation of the bare series at PBE, TPSS, B3LYP, and $\omega$B97X-D3 on the PBE0/ZORA geometries, with identical basis sets and without the RIJCOSX approximation. The adiabatic reduction energy of the U(IV)/U(III) couple is reproduced closely: at CN~5, where all four functionals converged, the values span $3.57$ to $3.95$~eV against $3.83$~eV at PBE0, and both the change of sign between CN~4 and CN~5 and the rise of roughly $11$~eV to CN~7 are recovered wherever the relevant species converged (B3LYP, $-1.66$~eV at CN~4 to $+9.29$~eV at CN~7). The preferred coordination numbers are similarly unchanged across functionals: CN~5 for U(IV), CN~4 for U(III), and CN~5 for Th(IV) in every case. The thorium series converged at all four functionals across the full CN~3-8 range, and its V-shaped profile is reproduced throughout, the CN~5 minimum lying $27.6$ to $41.6$~kJ~mol$^{-1}$ below CN~6 in every case. Full results are collected in the Supporting Information.

\section{Results}
\label{sec:Results}

\subsection{Comparison of U(III), Th(IV), and U(IV) Coordination}
\label{sec:ucomp}

The clearest structural difference between the three species is how the An-F bond reacts to oxidation state at fixed coordination number. Actinide contraction pulls the bond in on passing from Th(IV) to U(IV) (at six-coordination $\langle r(\mathrm{Th\text{-}F})\rangle = 2.250$~\AA{} falls to $\langle r(\mathrm{U\text{-}F})\rangle = 2.198$~\AA) but reducing U(IV) to U(III) reverses it, pushing the bond back out to $2.201$~\AA{} in $[\text{UF}_6]^{3-}$. The added $5f$ electron expands the coordination sphere, partly cancelling the contraction that the higher nuclear charge of uranium would otherwise force. This effect grows with coordination number; at CN~7 the U(III) bond ($2.283$~\AA) exceeds the U(IV) value ($2.150$~\AA) by more than $0.13$~\AA, though the U(IV) figure is itself depressed by the asymmetric, distorted geometry $[\text{UF}_7]^{3-}$ adopts (Section~\ref{sec:physbasis}). The electronic origin shows up in the metal spin population, which sorts the oxidation states cleanly and independently of coordination environment: $2.01$-$2.08$ unpaired electrons on uranium across the U(IV) series against $3.00$-$3.06$ across U(III), a separation of very nearly one electron reproduced by every functional tested (Supporting Information). The reduced $f$ populations separate thorium from uranium by about two electrons, $14.4$-$14.6$~$e$ for $5f^0$ Th(IV) against $16.0$-$17.0$~$e$ across the uranium series (Table~S6; these are total $4f+5f$ populations, the filled $4f^{14}$ semicore of the all-electron basis included, Section~\ref{sec:Compdetails}), but they do not resolve the two uranium oxidation states from one another, for the reason set out in Section~\ref{sec:physbasis}. We therefore identify the oxidation states by spin population throughout. That electron does little bonding work. The Mayer An-F bond order in the U(III) series falls steadily from $0.585$ at CN~4 to $0.418$ at CN~7 (Table~S4), following the dilution of metal-ligand overlap as fluorides are added, and lying below the corresponding U(IV) values at low coordination.
 
\begin{figure}[htbp]
\centering
\includegraphics[width=0.4\textwidth]{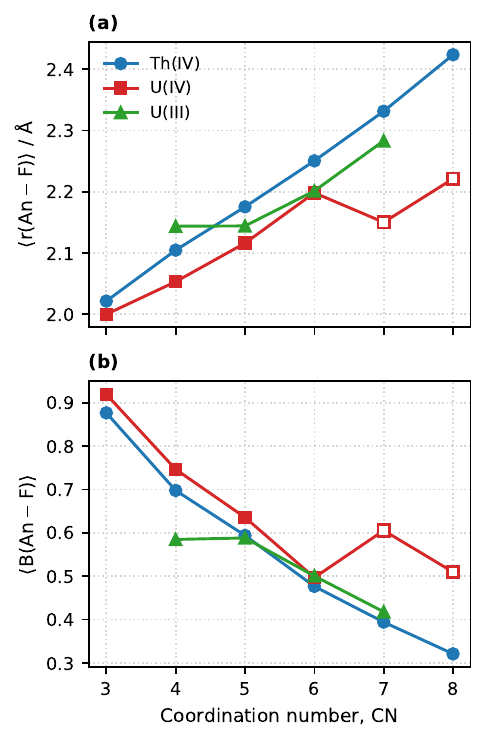}
\caption{(a) Mean An-F bond length and (b) Mayer bond order versus coordination number for the bare Th(IV), U(IV), and U(III) fluoride series. Open U(IV) markers at CN~7-8 are the contracted $[\text{UF}_7]^{3-}$/$[\text{UF}_8]^{4-}$ points discussed in Section~\ref{sec:physbasis}.}
\label{fig:struct_bonding}
\end{figure}
 
However, the clearest signal for redox comes from the frontier-orbital structure. The HOMO-LUMO gap of the U(III) complexes collapses monotonically as coordination rises, from $2.65$~eV in $[\text{UF}_4]^-$ through $0.76$ and $0.53$~eV, and at CN~7 it closes entirely; the gap evaluates to $-0.01$~eV, an inversion in which the highest occupied level lies marginally above the lowest virtual one and the Aufbau principle is violated (Table~S8). We do not read the sign as physically meaningful at this magnitude; what it establishes is that the frontier levels of $[\text{UF}_7]^{4-}$ are degenerate to within the resolution of the calculation, so that no gap remains to be reported. U(IV) narrows the same way, its gap reaching $0.53$~eV at CN~7 and $0.52$~eV at CN~8, and both uranium series sit far beneath the wide $2.7$-$9.0$~eV gaps of the closed-shell thorium complexes. This near-degeneracy of frontier levels in the highly coordinated uranium species follows from the dense $5f$ manifold pinned at the frontier, and it means electron addition or removal gets progressively easier as the fluoride shell grows. That is exactly the state in which the U(IV)/U(III) couple controls redox and corrosion chemistry in the melt, and we quantify it in Section~\ref{sec:redox}.
 
\begin{figure}[!htbp]
\centering
\includegraphics[width=0.8\linewidth]{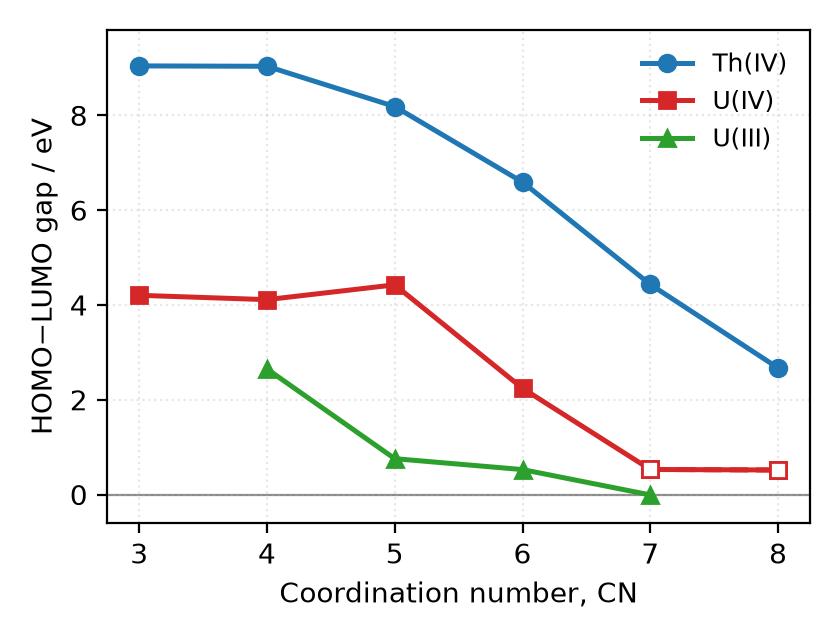}
\caption{HOMO-LUMO gap versus coordination number for the bare Th(IV), U(IV), and U(III)
fluoride complexes. The closed-shell Th(IV) gaps remain wide at every coordination number,
falling only from $9.0$ to $2.7$~eV across the series. Both uranium series lie far beneath them
and narrow as the fluoride shell grows: U(IV) reaches $0.53$~eV at CN~7, while the U(III) gap
closes entirely, evaluating to $-0.01$~eV at CN~7 where the highest occupied level lies
marginally above the lowest virtual one. Open U(IV) markers at CN~7-8 denote the contracted
$[\text{UF}_7]^{3-}$/$[\text{UF}_8]^{4-}$ species discussed in Section~\ref{sec:physbasis}.}
\label{fig:gap_vs_cn}
\end{figure}

\subsection{Preferred Coordination Numbers and the Effect of LiF Solvation}
\label{sec:coordpref}
 
\begin{table}[htbp]
\centering
\small
\caption{Relative Gibbs free energies $\Delta G_{\mathrm{rel}}$ (kJ\,mol$^{-1}$) of the bare
$[\text{AnF}_n]^{m-}$ complexes, referenced to the most stable CN of each oxidation state at each
temperature ($\Delta G_{\mathrm{rel}}\equiv 0$ at the minimum). Free-fluoride balancing with the
Sackur-Tetrode free energy as in the text. The $[\text{UF}_3]^+$ free energies at $973$ and
$1273$~K are withheld: its frequency calculation did not converge at $973$~K and returned a
second-order saddle at $1273$~K.}
\label{tab:relG}
\begin{tabular}{llrrr}
\hline\hline
Ox.\ state & CN & $298$~K & $973$~K & $1273$~K\\
\hline
Th(IV) & 3 & $1318.5$ & $1160.4$ & $1090.7$\\
       & 4 & $379.8$  & $311.1$  & $280.7$\\
       & \textbf{5} & \textbf{0.0} & \textbf{0.0} & \textbf{0.0}\\
       & 6 & $75.0$   & $162.6$  & $201.3$\\
       & 7 & $654.0$  & $823.2$  & $899.1$\\
       & 8 & $1597.8$ & $1814.9$ & $1910.2$\\
\hline
U(IV)  & 3 & $1376.8$ & -       & -\\
       & 4 & $428.6$  & $354.9$  & $323.0$\\
       & \textbf{5} & \textbf{0.0} & \textbf{0.0} & \textbf{0.0}\\
       & 6 & $95.4$   & $178.3$  & $214.7$\\
       & 7 & $687.1$  & $850.0$  & $921.5$\\
       & 8 & $1556.5$ & $1797.3$ & $1902.8$\\
\hline
U(III) & \textbf{4} & \textbf{0.0} & \textbf{0.0} & \textbf{0.0}\\
       & 5 & $124.6$  & $200.4$  & $233.4$\\
       & 6 & $487.3$  & $642.6$  & $710.1$\\
       & 7 & $1357.8$ & $1588.6$ & $1689.1$\\
\hline\hline
\end{tabular}
\end{table}

Left to itself in the gas phase, every one of the three actinides prefers to be undercoordinated. Referencing the relative free energies of Table~\ref{tab:relG} to the most stable species at each temperature, with stoichiometric balancing by free fluoride, Th(IV) and U(IV) both settle on the five-coordinate $[\text{AnF}_5]^-$ ion while U(III) settles on four-coordinate $[\text{UF}_4]^-$, and the preference holds at all three temperatures. For Th(IV) the next-nearest coordination number is CN~6, lying $75.0$, $162.6$, and $201.3$~kJ\,mol$^{-1}$ above CN~5 as the temperature climbs, and the ordering of all six thorium species follows the same way. As temperature increases, higher coordination is accordingly penalized; eight-coordinate $[\text{ThF}_8]^{4-}$ sits $1597.8$~kJ\,mol$^{-1}$ above the CN~5 reference at $298$~K and $1910.2$~kJ\,mol$^{-1}$ above it at $1273$~K, because each fluoride bound is translational and rotational entropy taken away from an otherwise free F$^-$. U(IV) follows similarly, with CN~6 lying $95.4$-$214.7$~kJ\,mol$^{-1}$ above CN~5 across the range with the high-CN species progressively destabilized. U(III) is the most severe of the three: CN~5 already costs $124.6$~kJ\,mol$^{-1}$ at $298$~K and $233.4$~kJ\,mol$^{-1}$ at $1273$~K.
 
Of course, this does not actually describe the coordination adopted in the salt. Experiment and molecular dynamics put the dominant thorium and uranium species at CN~6-8 \cite{bessada2017high,smith2019exafs,bessada2020investigation,smith2022structure,dai2015molecular,liu2014theoretical,guo2019theoretical}, well above the gas-phase numbers. This discrepancy comes from a lack of description of the surrounding molten salt environment. The dominant penalty for the bare ion is the formal negative charge piling up as fluorides are added, and in the melt that charge is screened by the counter-ions that is not correctly described in the gas-phase model. However, the microsolvated optimizations resolve this and identify the controlling variable directly. Tables~\ref{tab:solv4lif} and~\ref{tab:solv8lif} collect the coordination-number response of every converged cluster. Sorted by the formal charge of the bare parent ion rather than by the metal, the response is close to a step function. The two cationic complexes climb at every shell for which they converged: $[\text{ThF}_3]^+$ goes $3\rightarrow5$ with four LiF units and $3\rightarrow6$ with eight, and $[\text{UF}_3]^+$ goes $3\rightarrow6$. The two neutral complexes climb likewise, $[\text{ThF}_4]^0$ reaching CN~7 at the second shell and $[\text{UF}_4]^0$ reaching CN~6 at the first. The boundary can be seen around the singly-charged anions; $[\text{ThF}_5]^-$ and $[\text{UF}_4]^-$ hold their bare coordination at both shells, while $[\text{UF}_5]^-$ holds at $+4$\,LiF and climbs $5\rightarrow6$ only once the second shell is present. From $2-$ onward, further fluoride coordination stops altogether (all six complexes of charge $2-$ and $3-$ hold their bare coordination at both shells) and at $4-$ it reverses: $[\text{ThF}_8]^{4-}$ sheds a fluoride to CN~7 at both shells, and $[\text{UF}_7]^{4-}$ sheds to CN~6 at the second shell, the displaced fluoride in that case being taken up by the LiF cage itself.
Solvation response is therefore determined by the charge of the complex, not the identity or oxidation state of the metal. We refer to this below as the charge rule. The mechanism follows directly from the bare-ion energetics above: a cationic or neutral complex sitting in a LiF cage still gains on binding an additional fluoride, and the incremental fluoride affinity falls monotonically as the complex's own charge becomes more negative, until near $1-$ it crosses zero, and by $4-$ the anion-anion repulsion within the first shell outweighs the counter-ion stabilization the cluster can supply. 
There are two comparisons that demonstrate this quite clearly. At CN~4, $[\text{ThF}_4]^0$ climbs while $[\text{UF}_4]^-$ holds. Moreover, at CN~5 the redox pair splits, $[\text{UF}_5]^-$ climbing to CN~6 at the second shell while $[\text{UF}_5]^{2-}$ holds; so we have the same metal at the same coordination number, differing only by one unit of charge and one $5f$ electron, responding oppositely. 
Sorted by charge, the three actinides behave alike, and the charge range each series happens to span accounts for any appearance to the contrary: the U(III) complexes carry charges of $1-$ to $4-$ across the whole CN~4-7 range studied and so are entirely within the regime where coordination holds, whereas the Th(IV) and U(IV) series reach into the neutral and cationic regime at CN~4 and CN~3 and climb there. Each metal is represented on both sides of the boundary only to the extent that its accessible coordination numbers put it there. The Mayer bond orders of Section~\ref{sec:physbasis} record a genuine $5f$ covalent contribution to the U-F bond, and it is visible in the exchange barriers of Section~\ref{sec:kinetics}, but it is not what determines the solvation response. Structurally, the climb shows up as a tightening of the inner An-F shell as first-shell fluorides are drawn into bridging arrangements with the surrounding Li$^+$ centers. The inner-shell An$\cdots$Li separations settle between $2.90$ and $4.34$~\AA{} across the whole set, with second-shell cations beyond $6$~\AA, consistent both with the radial placement used to build the clusters and with the Li$^+$-F$^-$ separations reported for the melt \cite{smith2019exafs,liu2014theoretical,DEWAN2013322}. The $\langle r(\text{An}\cdots\text{Li})\rangle$ values of Table~\ref{tab:solv8lif} average over both shells and should be read with the individual distances in the Supporting Information; the four-LiF clusters that climbed carry the shortest contacts, $[\text{UF}_4]^0\!\cdot\!4$LiF averaging $3.375$~\AA{} over all four cations. The coordination shells themselves are clearly defined, as in every cluster the sorted An-F distances break by at least $1.78$~\AA{} between the coordinating fluorides and the next-nearest, so the coordination-number assignments of Tables~\ref{tab:solv4lif} and~\ref{tab:solv8lif} do not depend on the cutoff chosen. The single exception is $[\text{UF}_7]^{4-}\!\cdot\!8$LiF, discussed below.
 
Three clusters have structural caveats. $[\text{UF}_6]^{3-}\!\cdot\!8$LiF and $[\text{UF}_7]^{4-}\!\cdot\!8$LiF each ejected a LiF unit during relaxation, to $53.3$ and $75.8$~\AA{} respectively; both are effectively $\cdot7$LiF clusters and their reported An$\cdots$Li averages exclude the ejected unit. $[\text{UF}_7]^{4-}\!\cdot\!8$LiF additionally has two minima separated by $1.5$~eV, of which the displaced-start structure is reported. It is the one cluster whose coordination number is not fixed by the distance break alone: its seventh fluoride sits at $2.995$~\AA, only $0.64$~\AA{} inside the next contact. The Li coordination settles it. That fluoride's nearest Li lies at $1.807$~\AA, closer than that of any fluoride retained in the first sphere and matching the fully expelled fluoride of the same cluster, so it has passed from the actinide to the LiF cage and the cluster is assigned CN~6. It is therefore the second of the two $4-$ complexes to shed, alongside $[\text{ThF}_8]^{4-}$, which is where the charge rule predicts fluoride loss should occur. Displaced-start controls on the remaining clusters agree with the reported geometries to within $0.005$~\AA. Imaginary modes were found in two clusters of the solvated set, both U(III) at the second shell: $[\text{UF}_6]^{3-}\!\cdot\!8$LiF carries two, at $-7.1$ and $-4.9$~cm$^{-1}$, and $[\text{UF}_7]^{4-}\!\cdot\!8$LiF carries three, at $-10.4$, $-6.0$ and $-3.6$~cm$^{-1}$. All five are below $11$~cm$^{-1}$ in magnitude, within the range attributable to numerical noise in the integration grid for clusters of this size and charge, and all five belong to the two clusters already flagged above for LiF ejection. The remaining twenty-five solvated structures are clean minima.

\begin{table}[htbp]
\centering
\footnotesize
\setlength{\tabcolsep}{5pt}
\caption{Coordination-number change and mean distances (\AA) in the $+4$\,LiF microsolvated
fluoride complexes, ordered by the formal charge $q$ of the bare parent ion. The $\cdot4$LiF
suffix is omitted from the species labels. A dash denotes no change in coordination number.}
\label{tab:solv4lif}
\begin{tabular}{lcccc}
\hline\hline
Species & $q$ & CN change & $\langle r_{\text{An-F}}\rangle^{\dagger}$
& $\langle r_{\text{An}\cdots\text{Li}}\rangle^{\ddagger}$\\
\hline
$[\text{ThF}_3]^+$    & $+1$ & $3\to5$ & $2.201$ & $6.329$\\
$[\text{UF}_3]^+$     & $+1$ & $3\to6$ & $2.223$ & $4.307$\\
$[\text{ThF}_4]^0$    & $0$  & \multicolumn{3}{c}{not converged}\\
$[\text{UF}_4]^0$     & $0$  & $4\to6$ & $2.202$ & $3.375$\\
\hline
$[\text{ThF}_5]^-$    & $-1$ & -      & $2.191$ & $3.998$\\
$[\text{UF}_5]^-$     & $-1$ & -      & $2.140$ & $4.456$\\
$[\text{UF}_4]^-$     & $-1$ & -      & $2.189$ & $3.984$\\
\hline
$[\text{ThF}_6]^{2-}$ & $-2$ & -      & $2.250$ & $4.057$\\
$[\text{UF}_6]^{2-}$  & $-2$ & -      & $2.198$ & $4.011$\\
$[\text{UF}_5]^{2-}$  & $-2$ & -      & $2.245$ & $4.015$\\
$[\text{ThF}_7]^{3-}$ & $-3$ & -      & $2.318$ & $3.551$\\
$[\text{UF}_7]^{3-}$  & $-3$ & -      & $2.271$ & $3.492$\\
$[\text{UF}_6]^{3-}$  & $-3$ & -      & $2.285$ & $4.073$\\
$[\text{UF}_8]^{4-}$  & $-4$ & -      & $2.341$ & $3.229$\\
$[\text{UF}_7]^{4-}$  & $-4$ & -      & $2.282$ & $3.517$\\
\hline
$[\text{ThF}_8]^{4-}$ & $-4$ & $8\to7$ & $2.329$ & $4.441$\\
\hline\hline
\end{tabular}
\\[2pt]
\begin{minipage}{\linewidth}
{\footnotesize $^{\dagger}$Averaged over the coordinating fluorides only, the coordination shell
being terminated at the first break of $\geq0.5$~\AA{} in the sorted An-F distance list; every
cluster shows a break of at least $1.78$~\AA{} at that point.
$^{\ddagger}$Averaged over all four Li positions. The $[\text{ThF}_3]^+$ value is raised by two
Li that relax to outer-sphere positions during the coordination climb.}
\end{minipage}
\end{table}
 
\begin{table}[htbp]
\centering
\footnotesize
\setlength{\tabcolsep}{5pt}
\caption{Coordination-number change and mean distances (\AA) in the $+8$\,LiF microsolvated
fluoride complexes, ordered by the formal charge $q$ of the bare parent ion. The $\cdot8$LiF
suffix is omitted from the species labels.
$\langle r_{\text{An}\cdots\text{Li}}\rangle$ averages over all eight Li positions except where
noted; the inner-shell Li lie at $2.90$-$4.34$~\AA{} (individual values in SI). Four clusters
did not converge.}
\label{tab:solv8lif}
\begin{tabular}{lcccc}
\hline\hline
Species & $q$ & CN change & $\langle r_{\text{An-F}}\rangle^{\dagger}$
& $\langle r_{\text{An}\cdots\text{Li}}\rangle^{\dagger}$\\
\hline
$[\text{ThF}_3]^+$    & $+1$ & $3\to6$ & $2.257$ & $4.013$\\
$[\text{UF}_3]^+$     & $+1$ & \multicolumn{3}{c}{not converged}\\
$[\text{ThF}_4]^0$    & $0$  & $4\to7$ & $2.315$ & $4.152$\\
$[\text{UF}_4]^0$     & $0$  & \multicolumn{3}{c}{not converged}\\
\hline
$[\text{UF}_5]^-$     & $-1$ & $5\to6$ & $2.208$ & $4.481$\\
$[\text{ThF}_5]^-$    & $-1$ & -      & $2.194$ & $5.636$\\
$[\text{UF}_4]^-$     & $-1$ & \multicolumn{3}{c}{not converged}\\
\hline
$[\text{ThF}_6]^{2-}$ & $-2$ & -      & $2.252$ & $5.373$\\
$[\text{UF}_6]^{2-}$  & $-2$ & -      & $2.197$ & $5.677$\\
$[\text{UF}_5]^{2-}$  & $-2$ & -      & $2.287$ & $5.001$\\
$[\text{ThF}_7]^{3-}$ & $-3$ & -      & $2.310$ & $4.697$\\
$[\text{UF}_7]^{3-}$  & $-3$ & -      & $2.273$ & $4.972$\\
$[\text{UF}_6]^{3-}$  & $-3$ & -      & $2.224$ & $4.808^{\ddagger}$\\
$[\text{UF}_8]^{4-}$  & $-4$ & \multicolumn{3}{c}{not converged}\\
$[\text{UF}_7]^{4-}$  & $-4$ & $7\to6$ & $2.217$ & $4.803^{\ddagger,\S}$\\
\hline
$[\text{ThF}_8]^{4-}$ & $-4$ & $8\to7$ & $2.330$ & $5.284$\\
\hline\hline
\end{tabular}
\\[2pt]
\begin{minipage}{\linewidth}
{\footnotesize $^{\dagger}$Averaging conventions as in Table~\ref{tab:solv4lif}.
$^{\ddagger}$These two clusters each ejected an LiF unit during relaxation, to $53.3$ and
$75.8$~\AA{} respectively, and are effectively $\cdot7$LiF; the tabulated average excludes the
ejected unit.
$^{\S}$The seventh fluoride lies at $2.995$~\AA{} from uranium, separated from the first six by
$0.64$~\AA, the smallest shell break in the set. It is assigned to the second sphere on the Li
coordination: its nearest Li lies at $1.807$~\AA, closer than that of any retained bridging
fluoride ($1.885$-$2.009$~\AA) and matching the fully expelled fluoride of the same cluster
($4.677$~\AA{} from U, $1.785$~\AA{} from Li). For comparison, all seven fluorides of the
$+4$\,LiF analogue lie between $2.153$ and $2.505$~\AA.}
\end{minipage}
\end{table}

\begin{figure}[htbp]
\centering
\includegraphics[width=\linewidth]{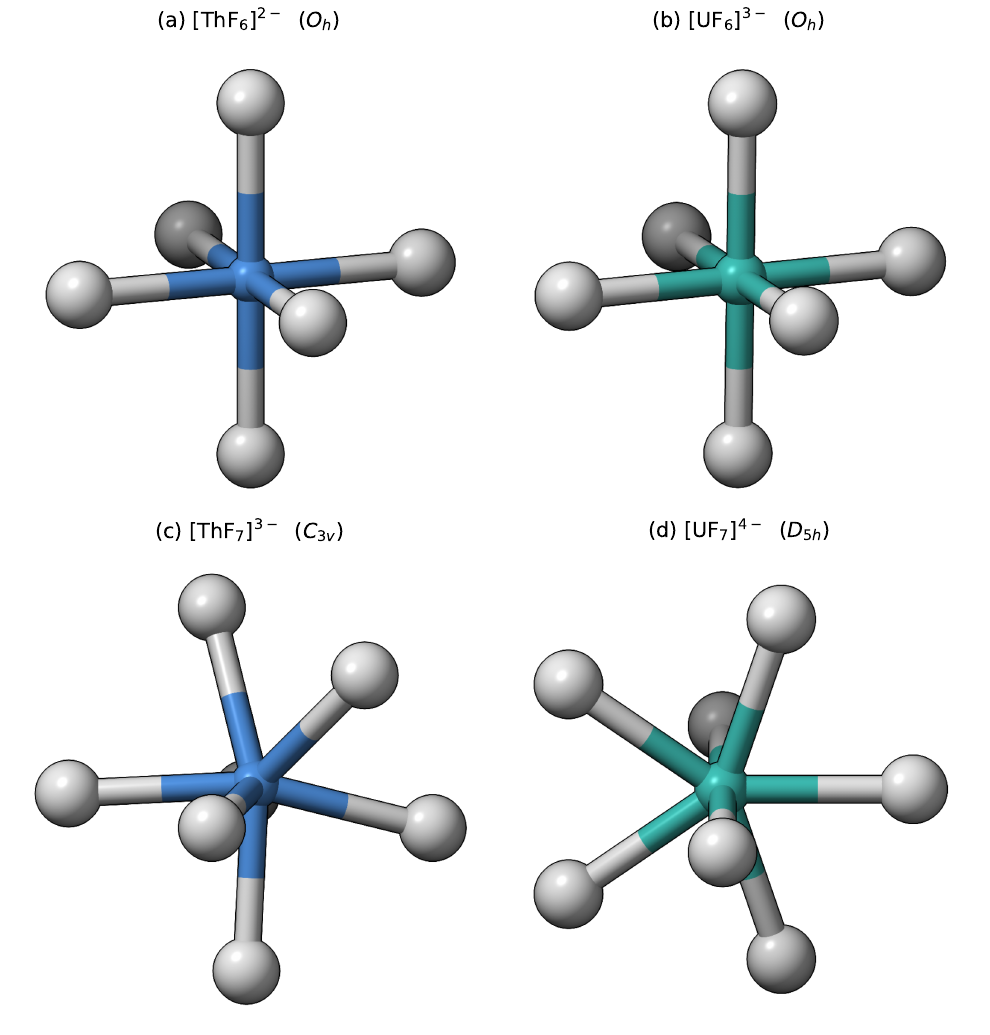}
\caption{Optimized coordination polyhedra of representative bare fluoride complexes:
(a) $[\text{ThF}_6]^{2-}$ and (b) $[\text{UF}_6]^{3-}$ at $\text{CN}=6$ (both $O_h$), and
(c) $[\text{ThF}_7]^{3-}$ ($C_{3v}$) and (d) $[\text{UF}_7]^{4-}$ ($D_{5h}$) at $\text{CN}=7$.
Metal centres are shown in blue (Th) and teal (U); fluoride in grey. An-F contacts are drawn for
internuclear distances below $2.6$~\AA{} (a visual cutoff for clarity only, not implying a
bonding criterion). Panels (a) and (b) are shown in the same orientation; the CN-6 octahedra are
essentially superimposable between Th(IV) and U(III), whereas the CN-7 polyhedra differ in
symmetry: a capped octahedron for Th(IV) versus a pentagonal bipyramid for U(III). All four
panels are drawn to a common scale.}
\label{fig:geometries}
\end{figure}

This answers the second question of Section~\ref{sec:Introduction}. The preferred bare coordination numbers are low (CN 5 for Th(IV) and U(IV), CN 4 for U(III)), and explicit LiF solvation raises them only for complexes on the cationic side of a boundary that falls at the singly-charged anions, moving those into or toward the experimental CN 6-8 window. The charge rule is what sets coordination number. The limitation that follows is specific to the anionic series. U(III) is experimentally reported near CN~$\approx6$ in fluoride melts, and none of our four U(III) clusters climbs at either shell, so for that series the finite cluster recovers neither the magnitude nor the direction of the solvation effect. This actually follows from the charge rule: the U(III) complexes are anionic at every coordination number sampled, and eight explicit LiF units supply too little counter-charge to overcome the fluoride-fluoride repulsion, where the full Madelung field of the bulk melt would. 

\begin{figure}[htbp]
\centering
\includegraphics[width=0.8\linewidth]{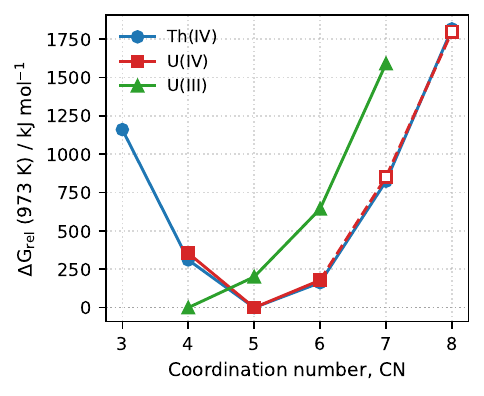}
\caption{Relative Gibbs free energy $\Delta G_\text{rel}$ at $973$~K versus coordination number for the bare Th(IV), U(IV), and U(III) fluoride series, referenced to the most stable CN of each series ($\Delta G_\text{rel}\equiv0$ at the minimum). Species of different CN are compared on a common basis by free-fluoride balancing, $[\text{AnF}_n]+(N n)\,\text{F}^-$, with the free-fluoride free energy from the Sackur-Tetrode equation. All three series are V-shaped, with minima at $\text{CN}=5$ for Th(IV) and U(IV) and $\text{CN}=4$ for U(III). Open U(IV) markers at $\text{CN}=7$-$8$ represent the contracted $[\text{UF}_7]^{3-}$/$[\text{UF}_8]^{4-}$ species discussed in Section~\ref{sec:physbasis}; the U(IV) $\text{CN}=3$ point is omitted as its $973$~K free energy was not available (frequency calculation unconverged). These are gas-phase (bare-ion) relative stabilities; the microsolvated melt favors higher CN through counter-ion screening, selectively by complex charge (see text).}
\label{fig:dG_vs_cn}
\end{figure}

\subsection{Physical Basis for Coordination Preferences}
\label{sec:physbasis}
 
The structural and thermodynamic trends all come from the same electronic origin, and we sort it into three factors. The first is the character of the An-F bond itself. Across every series the Mayer bond order falls monotonically as fluorides are added (from $0.877$ in $[\text{ThF}_3]^+$ to $0.321$ in $[\text{ThF}_8]^{4-}$, and from $0.919$ in $[\text{UF}_3]^+$ to comparably low values at high coordination (Table~S4)) which is what one expects when a fixed metal valence contribution is shared among more and more ligands. But at matched coordination the uranium bond orders exceed the thorium ones ($0.747$ for $[\text{UF}_4]^0$ against $0.698$ for $[\text{ThF}_4]^0$); that gap is a $5f$ covalent contribution to U-F bonding, absent for the $5f^0$ thorium center whose bonding is essentially ionic. This covalency binds the U(IV) first shell more tightly than thorium's and is visible in the exchange barriers of Section~\ref{sec:kinetics}. The second factor is the anionic charge accumulating as the shell grows, and it is the one that carries through to the solvated clusters. The reduced orbital populations of Table~S6 separate thorium from uranium clearly: the actinide $f$ population (total $4f+5f$, Section~\ref{sec:Compdetails}) sits at $14.4$-$14.6$~$e$ for $5f^0$ Th(IV) and rises to $16.0$-$17.0$~$e$ across the uranium series, the $5f$ occupation accounting for the whole of that difference. Between the two uranium oxidation states the same analysis is not decisive. At matched coordination the U(III) $f$ population exceeds the U(IV) one by $+0.41$~$e$ at CN~4 and $+0.35$~$e$ at CN~7, but by only $+0.07$~$e$ at CN~5 and $-0.003$~$e$ at CN~6, with a compensating gain of up to $1.0$~$e$ appearing in the $s$ channel instead. That $s$ gain is not a $7s$ occupation but the same diffuse-density misassignment that produces the unphysical L\"owdin charges discussed below, since it appears only in the uranium species carrying charges of $2-$ and beyond ($s = 13.04$-$13.06$~$e$, against $12.02$-$12.10$~$e$ at lower charge, and including the U(IV) $[\text{UF}_7]^{3-}$ and $[\text{UF}_8]^{4-}$ complexes), and it is absent from the thorium series at every charge, $[\text{ThF}_8]^{4-}$ included. The reduced populations therefore establish the $5f^0$/$5f^n$ distinction between the metals but not the $f^2$/$f^3$ distinction within uranium, for which we use the spin population. The $p$ and $d$ populations, by contrast, barely move across all species. The bare ion has no counter-charge to stabilize the charge it accumulates, so high coordination is strongly disfavored in the gas phase; the explicit Li$^+$ counter-ions of the microsolvated clusters offset the build-up. The third factor is where the $5f$ manifold sits relative to the frontier. For the closed-shell Th(IV) complexes the HOMO-LUMO gap is wide at every coordination number, from $9.0$~eV at CN~3 down to $2.7$~eV at CN~8 (Table~S8), so the thorium geometries are electronically rigid and their relative stability follows almost entirely from the electrostatics and entropy already discussed. The uranium complexes are qualitatively different, as their gaps lie much further below the thorium values at every coordination number and collapse as the shell grows (U(III) from $2.65$~eV in $[\text{UF}_4]^-$ to a closed, marginally inverted gap of $-0.01$~eV at CN~7, U(IV) reaching comparably small values at high coordination) because the partially occupied $5f$ manifold sits at the frontier and is driven toward degeneracy as the ligand field spreads over more fluorides. This has two consequences: it makes the high-coordinate uranium geometries electronically soft and prone to multireference character, and it drives the geometric distortions visible in the optimized structures, i.e. the asymmetric arrangements adopted by the seven- and eight-coordinate uranium fluorides.
 
Two of those high-CN U(IV) species initially appear to break the trends. $[\text{UF}_7]^{3-}$ and $[\text{UF}_8]^{4-}$ show short An-F bonds, negative L\"owdin charges on the metal, and out-of-trend Mayer bond orders (Tables~S4,S5). These charges are an artifact of the L\"owdin orthogonalization as already discussed in the Methods. The Mulliken charges for the same species are positive and sensible ($+1.57$~$e$ and $+1.67$~$e$), and the spin populations ($\approx 2$ unpaired electrons) and $\langle S^2\rangle$ values (clean $f^2$ triplets) confirm the expected U(IV) electronic structure, so we report Mulliken charges for these two complexes.
The contracted An-F distance of $[\text{UF}_7]^{3-}$ ($2.150$~\AA), which falls below the otherwise monotonic rise of bond length with coordination number, is a different matter, and we interpret it as real. Re-optimization from an expanded starting geometry (initial U-F~$\approx 2.30$~\AA) relaxes to a distinct, less-contracted structure ($\langle r(\text{U-F})\rangle = 2.28$~\AA) lying $5.9$~kcal\,mol$^{-1}$ higher in energy, so among the geometries examined the contracted one is the preferred minimum. The elevated Mayer bond order that accompanies it ($0.605$, against the CN~6 value of $0.497$) follows directly from the contraction. Across the U(IV) series the bond order tracks the An-F distance closely (Pearson $r = -0.98$); the line fit to the other five species predicts $0.6021$ at $2.150$~\AA{} against an observed $0.6054$, so $[\text{UF}_7]^{3-}$ falls on the relation to within $0.6\%$. The out-of-trend appearance comes about only because the bond is shorter than that of $[\text{UF}_6]^{2-}$, not from any difference in electronic structure. $[\text{UF}_8]^{4-}$, for its part, behaves as expected ($2.221$~\AA, longer than $[\text{UF}_6]^{2-}$).
 
These three factors answer the third question of Section~\ref{sec:Introduction}. The relative stability of the coordination geometries across the actinide series is set by a fixed, weakly covalent An-F interaction that the $5f$ manifold modulates (adding covalency for uranium that thorium lacks) combined with the metal's role in absorbing the anionic charge that accumulates with coordination, and set by the energetic position of the $5f$ levels. The last factor separates the oxidation states most clearly. The closed-shell, large-gap thorium geometries are stable and rigid; the open-shell uranium geometries grow progressively softer and more easily oxidized or reduced as the fluoride shell expands.

\subsection{Redox Energetics}
\label{sec:redox}

\begin{table}[htbp]
\centering
\small
\caption{Adiabatic reduction energies for the U(IV)/U(III) couple in the bare
$[\text{UF}_n]^{m-}$ complexes. $\Delta E_\mathrm{adiab}$ and $\Delta G$ are defined as
$E/G[\text{U(III)}] - E/G[\text{U(IV)}]$ for U(IV) $+\,e^- \rightarrow$ U(III); negative values
indicate that U(III) is the more stable state at that coordination number. These are intrinsic
reduction energies, reported without referencing to an absolute electrode.}
\label{tab:redox}
\begin{tabular}{lrrr}
\hline\hline
CN & $\Delta E_\mathrm{adiab}$ / eV & $\Delta G_{298\,\mathrm{K}}$ / eV & $\Delta G_{973\,\mathrm{K}}$ / eV\\
\hline
4 & $-1.51$ & $-1.94$ & $-2.00$\\
5 & $3.83$  & $3.79$  & $3.75$\\
6 & $6.62$  & $6.56$  & $6.49$\\
7 & $9.53$  & $9.45$  & $9.33$\\
\hline\hline
\end{tabular}
\\[2pt]
{\footnotesize Solvated reduction energies are not reported; see text. Inner-sphere
reorganization energies from Equation~\ref{eq:lambda} are given in the Supporting Information
(Table~S16) and are discussed in the text.}
\end{table}

The frontier-orbital collapse of Section~\ref{sec:physbasis} has a thermodynamic counterpart. Table~\ref{tab:redox} gives the adiabatic reduction energy $\Delta E_\mathrm{adiab} = E[\text{U(III)}_\mathrm{relaxed}] - E[\text{U(IV)}_\mathrm{relaxed}]$ for the half-reaction U(IV) $+\,e^- \rightarrow$ U(III) against coordination number. These are intrinsic reduction energies rather than electrode potentials on the electrochemical scale. They are reported as computed, without referencing to an absolute electrode (Section~\ref{sec:Compdetails}), so it is their variation with coordination rather than their magnitudes that affects the physics of the melt.

These vary enormously. Across the bare series $\Delta E_\mathrm{adiab}$ climbs from $-1.51$~eV at CN~4 through $+3.83$ and $+6.62$~eV to $+9.53$~eV at CN~7: a swing of roughly $11$~eV over four coordination numbers, changing sign between CN~4 and CN~5. At low coordination the reduction is exergonic, U(III) is the stable state, and the complex is reducing; at high coordination the reduction turns strongly endergonic, U(IV) is favored, and the complex is oxidizing. Coordination number is therefore a redox control variable in its own right, and it follows the frontier-gap behavior, since the U(III) gap collapses from $2.65$~eV at CN~4 to closure at CN~7 over exactly the range where the reduction energy swings positive, both following the progressive destabilization of the added $5f$ electron as the ligand field spreads. Inner-sphere reorganization energies for the bare couple, from the four-point construction of Equation~\ref{eq:lambda}, are collected in the Supporting Information (Table~S16). We report them as an observation rather than a result, and in fact as a negative one. The four values obtained are $+0.058$, $-0.352$, $-0.420$ and $+0.251$~eV at CN~4 through CN~7. A reorganization energy is non-negative by construction, so three of the four are unphysical, and the four-point construction has clearly not returned a meaningful decomposition here. Two features of the calculation account for it: the cross-geometry points are single-determinant descriptions of states whose single-reference character is itself in question (Section~\ref{sec:multiref}), and the numerical integration of the density in the most highly charged of these anions is not tight enough to support differences of this size, which are of order a few tenths of an eV between total energies of order $3\times10^{4}$~$E_\mathrm{h}$. We therefore draw no conclusion from $\lambda_i$ and report the values only to document that the construction failed.

We do not report solvated reduction energetics, for two reasons that apply over different ranges of coordination number.
The first is a collapse of the electronic state. Table~\ref{tab:spinpop} gives the uranium Mulliken spin population in the microsolvated U(III) clusters. At CN~4 and CN~5 the formal $5f^3$ configuration is retained cleanly and is essentially unaffected by solvation: $3.102$ at CN~4 and $3.082$ at CN~5 with four LiF units, and $3.076$ at CN~5 with eight. From CN~6 the third electron is lost to the fluoride and LiF environment. The loss is not progressive across coordination number but a threshold sitting between CN~5 and CN~6; and it is completed by the second shell rather than by the first. With four LiF units the collapse is partial and not even monotonic, giving intermediate populations of $2.345$ at CN~6 and $2.602$ at CN~7, the higher coordination number retaining more spin than the lower. With eight the collapse is complete and flat, $2.080$ at CN~6 and $2.081$ at CN~7, a clean ligand-reduced $5f^2$ state. Two conditions are therefore jointly required: a coordination number of six or more, and a second solvation shell.
The metal charge substantiates this independently. The clusters that retain $5f^3$ carry uranium Mulliken charges between $+1.76$ and $+1.95$, while the two fully collapsed clusters carry the highest charges of the set, $+2.43$ at CN~6 and $+2.28$ at CN~7 with eight LiF units. The electron lost from the spin density reappears as charge transferred off the metal, which is what a ligand-reduced state requires and what a mere spin-coupling artifact would not produce.

\begin{table}[htbp]
\centering
\small
\caption{Uranium Mulliken spin population and atomic charge in the microsolvated U(III) clusters, by coordination number and solvation shell. The formal $5f^3$ configuration is retained at CN~4 and CN~5 and lost from CN~6, completely so once the second shell is present. The metal charge rises as the spin falls, locating the third electron on the ligand environment.}
\label{tab:spinpop}
\begin{tabular}{lcccc}
\hline\hline
 & \multicolumn{2}{c}{Spin population} & \multicolumn{2}{c}{Mulliken charge / $e$}\\
CN & $+4$\,LiF & $+8$\,LiF & $+4$\,LiF & $+8$\,LiF\\
\hline
4 & $3.102$ & -      & $+1.905$ & -\\
5 & $3.082$ & $3.076$ & $+1.756$ & $+1.950$\\
6 & $2.345$ & $2.080$ & $+2.128$ & $+2.434$\\
7 & $2.602$ & $2.081$ & $+1.906$ & $+2.278$\\
\hline\hline
\end{tabular}
\\[2pt]
{\footnotesize Uranium Mulliken spin populations and atomic charges from the PBE0/ZORA property single points. The CN~4 $+8$\,LiF cluster was not obtained, so that entry is empty. Note the non-monotonicity at the first shell, where CN~7 retains more spin density than CN~6. The CN~6 $+8$\,LiF entry is structure-sensitive: a second minimum of that cluster gives $2.385$, an intermediate value, against $2.080$ for the structure reported here.}
\end{table}
 
Whether the $f^2$ solution is an artifact of the single determinant rather than a physical reduction is taken up in Section~\ref{sec:multiref}, where the active-space treatment proves inconclusive. For the present purpose the consequence is narrower and does not depend on it: at CN~6 and CN~7 the solvated U(III) endpoint is not the state whose energy we want, so a reduction energy computed from it would be meaningless whichever way that question is eventually settled. At CN~4 and CN~5 this objection does not apply.
 
The second reason does apply at CN~5, and it is in fact structural. A redox pair differs by one unit of charge by construction ($[\text{UF}_n]^{(n-4)-}$ against $[\text{UF}_n]^{(n-3)-}$ at matched $n$) and charge is precisely what sets the solvation response (Section~\ref{sec:coordpref}). The two members of any pair therefore sit at adjacent points on the charge-response curve and do not need to keep the same coordination number when relaxed in the same LiF environment. They do not. At CN~5 with eight LiF units the U(IV) member $[\text{UF}_5]^-$ climbs to CN~6 while the U(III) member $[\text{UF}_5]^{2-}$ holds at CN~5, so the pair prepared at matched coordination does not remain matched through relaxation. This is the charge rule operating exactly as the rest of the section describes rather than a calculation failure. Constraining both sides would force the match, but at a coordination number one of the two species does not adopt.
The lability behind this follows the kinetics of Section~\ref{sec:kinetics} appearing directly in the structural optimizations. Measured exchange barriers span roughly $4$-$17$~kcal\,mol$^{-1}$, which is little barrier at all on the potential-energy surface an optimizer explores, let alone at $973$~K. The coordination sphere of a microsolvated actinide is labile by construction, which means coordination number is not a fixed property one can assign to a solvated redox pair and hold fixed while the electron is added. It is a time-averaged quantity, which is why the experimental values are non-integer ($\langle \mathrm{CN}\rangle = 7.42$-$7.92$ for U(IV) \cite{bessada2020investigation,smith2022structure,ocadiz2021new}) and why our bare complexes, having no shell to exchange with, can hold integer coordination while the solvated ones cannot. We therefore report the reduction energetics for the bare series, where the comparison is well posed. The two obstacles in the solvated case are independent and their ranges are different: from CN~6 the electronic state itself is wrong at the single-determinant level, and at every coordination number the charge asymmetry intrinsic to a redox pair prevents the two sides from being held at matched coordination. Whether the collapsed state is correctly described by any single determinant is a separate question, which the active-space calculations of Section~\ref{sec:multiref} were unable to answer.

\subsection{Fluoride-Exchange Kinetics}
\label{sec:kinetics}
 
\begin{table*}[t]
\centering
\small
\caption{Fluoride-exchange barriers from transition-state optimization, referenced to the relaxed-scan reactant. The low-coordination additions ($3\rightarrow4$, $4\rightarrow5$, $5\rightarrow6$) are barrierless and are omitted here; full scan data are in the Supporting Information.}
\label{tab:kinetics}
\begin{tabular}{ll r r}
\hline\hline
Ox.\ state & Reaction & $\Delta E^{\ddagger}$ / kcal\,mol$^{-1}$ & $\nu_\mathrm{im}$ / cm$^{-1}$\\
\hline
Th(IV) & $[\text{ThF}_6]^{2-}\rightarrow[\text{ThF}_7]^{3-}$ & $16.63$ & $-109.9$\\
       & $[\text{ThF}_7]^{3-}\rightarrow[\text{ThF}_8]^{4-}$ & $4.58$  & $-132.0$\\
\hline
U(IV)  & $[\text{UF}_6]^{2-}\rightarrow[\text{UF}_7]^{3-}$   & $13.86$ & $-120.7$\\
       & $[\text{UF}_7]^{3-}\rightarrow[\text{UF}_8]^{4-}$   & $4.44$$^{a}$ & -\\
\hline
U(III) & $[\text{UF}_6]^{3-}\rightarrow[\text{UF}_7]^{4-}$   & $7.01$  & $-136.8$\\
\hline\hline
\end{tabular}
\\[2pt]
{\footnotesize Each transition state carries a single imaginary mode whose eigenvector is a single-fluoride departure along the An-F coordinate, with that fluoride carrying $0.97$-$0.98$ of the normalized displacement, and a metal spin population matching the formal $5f$ occupation ($0.00$ for Th(IV), $2.07$ and $2.04$ for U(IV), $3.05$ for U(III)). $^{a}$Relaxed-scan bound: the saddle search for this step converged to a delocalized stationary point below the scan maximum and is not the exchange transition state (Supporting Information).}
\end{table*}

Table~\ref{tab:kinetics} collects the barriers from transition-state optimization on the relaxed An-F scans. Twelve of the thirteen scans ran to completion, and four of the five transition states converged to a first-order saddle carrying a single imaginary mode whose eigenvector is a single-fluoride departure along the An-F coordinate and whose spin state matches the formal $5f$ occupation. The $[\text{UF}_7]^{3-} \rightarrow [\text{UF}_8]^{4-}$ saddle search converged to a delocalized stationary point below the scan maximum; for that step we report the relaxed-scan bound and exclude the located structure (Supporting Information). The barriers are uniformly low, spanning roughly $4$ to $17$~kcal\,mol$^{-1}$, with the lowest-coordination fluoride additions effectively barrierless. Exchange is practically effortless across the reactor-relevant range, and the salt should reach coordination and redox equilibrium quickly on operational timescales. Two trends organize them. First, the barriers fall with coordination number: the $7\rightarrow8$ additions cost only ${\sim}4.5$~kcal\,mol$^{-1}$ against ${\sim}7$-$17$~kcal\,mol$^{-1}$ for the $6\rightarrow7$ additions, i.e. adding fluoride to an already-crowded shell is easier than reaching that shell in the first place. Second, and more interestingly, the barriers are systematically lowest for U(III) and highest for Th(IV) at matched coordination: for the $6\rightarrow7$ step they rise from $7.01$~kcal\,mol$^{-1}$ for U(III) to $13.86$ for U(IV) and $16.63$ for Th(IV). That ordering is the kinetic signature of the electronic structure driving the redox behavior. The reduced, small-gap U(III) center is electronically the softest of the three, and its softness lowers the barrier to rearranging the fluoride shell; the closed-shell, wide-gap Th(IV) center is the most rigid and the slowest to exchange. These barriers are not an abstraction, and the solvated optimizations show them at work. Every cluster that changed coordination number did so during ordinary geometry relaxation, recruiting fluoride from the surrounding shell or losing it without any imposed reaction coordinate: $[\text{ThF}_3]^+$ climbing $3\to5$ and $3\to6$, $[\text{UF}_3]^+$ $3\to6$, $[\text{UF}_4]^0$ $4\to6$, $[\text{ThF}_4]^0$ $4\to7$, $[\text{UF}_5]^-$ $5\to6$, and $[\text{ThF}_8]^{4-}$ and $[\text{UF}_7]^{4-}$ shedding to CN~7 and CN~6 respectively. Barriers of a few kcal\,mol$^{-1}$ are what makes that possible. It follows that the clusters which held their coordination did not do so because a barrier prevented it (the barriers are the same order throughout) but because the addition of another fluoride ligand is thermodynamically unfavorable at their charge.

\subsection{Active-Space Analysis of the Collapsed States}
\label{sec:multiref}
 
The state collapse of Section~\ref{sec:redox} raises a question the single-determinant calculations cannot answer: is the $f^2$ state that PBE0/ZORA converges to a physically real reduction of U(III), or an artifact of forcing a single determinant onto a state that has none? Addressing it requires an active-space treatment, and the attempt is reported here in full including its failure. Active spaces were selected by atomic-valence character rather than by orbital energy, using an atomic-valence active-space (AVAS) projection onto the uranium valence-$f$ shell, evaluated in an ANO-RCC minimal basis on the $+8$LiF cluster geometries. At CN~5 and CN~6 the construction behaves as intended at the default threshold of $0.25$. It returns seven orbitals with $5f$ projection weights of $0.905$-$0.978$ at CN~5 and $0.851$-$0.997$ at CN~6, separated from the highest rejected orbital by a wide margin in both cases ($0.095$ and $0.149$ respectively), so the selection is unambiguous and a CAS(3,7) follows without further intervention.
 
At CN~7 the same setup fails. At threshold $0.25$ only a single orbital of the uranium $5f$ shell (projection weight $0.996$) is occupied in the reference determinant; the remaining six $5f$-derived orbitals are virtual. AVAS accordingly selects seven orbitals but only one electron, fewer than the imposed multiplicity requires, and no CAS(3,7) exists to be built. The $5f$ manifold at CN~7 is not just pushed toward the frontier, as the gap data of Section~\ref{sec:physbasis} indicate; in the reference determinant it is very largely emptied. That is the orbital-space expression of the same ligand-reduced collapse the spin populations and metal charges of Section~\ref{sec:redox} record, arrived at independently, and it does not depend on any subsequent correlated calculation.
Obtaining a usable space at CN~7 therefore required lowering the selection threshold to $0.10$, which yields CAS(5,9). This space is not comparable in quality to the other two. Seven of its nine orbitals carry $5f$ projection weights between $0.847$ and $0.996$, comparable to the CN~5 and CN~6 sets, but the remaining two carry weights of only $0.115$ and $0.152$ and are therefore approximately $85$-$89\%$ ligand in character. They are admitted only because the threshold was lowered past them, and they supply the two additional electrons that distinguish CAS(5,9) from CAS(3,7). The cut between accepted and rejected orbitals is itself clean, the highest rejected orbital lying at $0.004$; what distinguishes this space from the other two is that the accepted set is not homogeneous, splitting into a $5f$-like group of seven and a ligand-like pair well below it.
\begin{table}[htbp]
\centering
\footnotesize
\setlength{\tabcolsep}{4pt}
\caption{AVAS-selected active spaces and CASCI expansions for the solvated U(III) clusters on the $+8$LiF geometries. Weights are projections onto the uranium valence-$f$ shell. The CN~7 space is not comparable to the other two:  it is obtained at a lowered threshold and its orbital set is not homogeneous in $5f$ character.}
\label{tab:avas}
\begin{tabular}{ccll}
\hline\hline
CN & Space & $5f$ weights & Leading config\\
\hline
5 & CAS(3,7)       & $0.905$-$0.978$     & $0.588$ \ (2)\\
6 & CAS(3,7)       & $0.851$-$0.997$     & $0.656$ \ (4)\\
7 & CAS(5,9)$^{a}$ & $0.847$-$0.996^{b}$ & $0.330/0.326^{c}$ \ (4)\\
\hline\hline
\end{tabular}
\\[2pt]
\begin{minipage}{\linewidth}
{\footnotesize
$^{a}$At a projection threshold of $0.10$. At the default $0.25$ no CAS(3,7) is constructible: AVAS selects seven orbitals but only one electron, six of the $5f$-derived orbitals being virtual in the reference determinant. $^{b}$Seven orbitals span this range; the space contains two further orbitals at $0.115$ and $0.152$, predominantly ligand in character, which enter only because the threshold was lowered. The highest rejected orbital lies at $0.004$, so the cut itself is clean; it is the accepted set that is not homogeneous. The highest rejected orbitals at CN~5 and CN~6 lie at $0.095$ and $0.149$. $^{c}$Near-degenerate leading pair. Determinant weights were printed to a threshold of $0.01$ and are counted above $0.05$. All CASCI calculations were run in the $M_S=3/2$ block, so the three-unpaired-electron configuration is imposed by the input rather than selected by the calculation; no $f^2$ (doublet) comparison was performed. Orbital-optimized CASSCF did not converge in any of eight attempts (text).}
\end{minipage}
\end{table}
 
Table~\ref{tab:avas} collects the CASCI expansions obtained in these spaces. The leading configuration carries a weight of $0.588$ at CN~5, $0.656$ at CN~6, and at CN~7 the two leading configurations are near-degenerate at $0.330$ and $0.326$. Taken at face value these would indicate substantial multiconfigurational character throughout and something close to a two-configuration state at CN~7. They cannot be taken at face value for two reasons.
 
The first is that the CASCI natural occupations are strongly fractional in the AVAS orbitals but become integral as soon as those orbitals are allowed to relax. In the AVAS basis the CN~5 occupations are $0.978$, $0.958$, $0.605$, $0.336$, $0.060$, $0.059$, $0.003$, and the CN~6 and CN~7 spaces are comparably fractional. Orbital-optimized CASSCF in the same spaces pushes every one of them to integers: $1,1,1,0,0,0,0$ at CN~5 and CN~6, and $2,1,1,1,0,0,0,0,0$ at CN~7. This occurs in all eight runs, at all three coordination numbers, and under every damping setting tested. An active space whose optimized occupations are integral holds a single determinant and carries no correlation; the fractional occupations (and with them the multiconfigurational appearance of the CASCI expansions) are properties of the unoptimized AVAS orbitals rather than of the states. The second is specific to CN~7. The near-degeneracy there arises in a space containing two orbitals that are predominantly ligand in character, and the leading configurations differ in the occupation of exactly those orbitals; the doubly occupied orbital appearing in both is one of them. A near-degenerate pair of this kind is what metal-versus-ligand mixing produces inside a contaminated active space, and it is not evidence of static correlation in the uranium $5f$ manifold. We therefore do not report the CN~7 expansion as a measure of multireference character.
 
Unfortunately the CASSCF calculations do not rescue the analysis, and their behavior is consistent with an ill-posed active space rather than with strong correlation. Eight runs were performed, all capped at $200$ macroiterations, all seeded from converged CASCI orbitals, and all terminated on convergence failure. Without damping the energy leaves the variational regime entirely and changes sign, rising from $-30385.281$ to $+60547.90$~$E_\mathrm{h}$ at CN~5, from $-30476.466$ to $+58842.96$~$E_\mathrm{h}$ at CN~6, and from $-30569.673$ to $+63547.75$~$E_\mathrm{h}$ at CN~7, with intermediate excursions above $+7\times10^{4}$~$E_\mathrm{h}$. A level shift of $2.0$~$E_\mathrm{h}$ with maximum rotation $0.03$ reduces but does not prevent the sign change. A shift of $3.0$~$E_\mathrm{h}$ with maximum rotation $0.015$ suppresses it: at CN~5 and CN~6 the energy rises by $2.4\times10^{2}$ and $1.6\times10^{3}$~$E_\mathrm{h}$ but remains negative throughout, and at CN~7 it descends instead, reaching $-30579.063$~$E_\mathrm{h}$, some $9.4$~$E_\mathrm{h}$ below the CASCI reference, and flattening to $\mathrm{d}E \sim 10^{-10}$~$E_\mathrm{h}$ over the final macroiterations. That plateau is not convergence, however; the orbital gradient at termination is $\lVert g \rVert = 9.67$ with a largest component of $-3.55$, so the calculation is stationary in energy only because the rotation cap prevents it from moving, and it remains far from a stationary point in the orbitals. We therefore report the following. At CN~5 and CN~6 a clean $5f$ valence active space can be constructed and a CAS(3,7) CASCI performed in it, but the resulting expansion is not a reliable measure of correlation because the orbitals are not optimized and every attempt to optimize them collapses the space to a single determinant. At CN~7 no $5f$ valence space exists at the standard threshold at all, and the space obtainable at a lowered threshold is ligand-contaminated to a degree that makes its expansion uninterpretable. Across all three, the active space as constructed is unstable under orbital optimization. We do not, on this evidence, characterize the strength of the correlation in these states.
 
There are two further limitations in this multireference treatment. First, the CASCI calculations were run in the $M_S = 3/2$ block throughout, so the three-unpaired-electron configuration is imposed by the input; no $f^2$ calculation was performed in these active spaces, and the recovery of a metal-localized $f^3$ solution is therefore a consequence of the multiplicity chosen rather than the outcome of a comparison between the two configurations. Separately, the AVAS projection is evaluated on the DFT reference determinant, which at CN~6 and CN~7 is the collapsed $f^2$ state; the extent to which the resulting selection is conditioned on that reference has not been established. What follows independently of all of this is the CN~7 selection failure. It requires no correlated calculation and no active space: at the standard threshold, six of the seven $5f$-derived orbitals of the solvated CN~7 cluster lie above the Fermi level of the reference determinant. That is a direct statement about where the $5f$ manifold sits, it agrees with the spin populations and metal charges of Section~\ref{sec:redox} and with the frontier-gap collapse of Section~\ref{sec:physbasis}, and it is the strongest evidence in this work that the third electron has genuinely left the metal in the highly coordinated solvated clusters.

\section{Discussion}
\label{sec:Discussion}

The coordination number of a dissolved actinide is not determined by the metal. It is imposed by the surroundings, and once fixed it propagates into the electronic structure, the redox thermodynamics, and the exchange kinetics. Everything else here follows from that. The bare-ion calculations make the point by contradiction: strip away the salt environment and each center settles at a low coordination number, nowhere near the CN~6-8 range that EXAFS and molecular dynamics establish for these ions in the melt \cite{bessada2017high,smith2019exafs,bessada2020investigation,smith2022structure}. Put the alkali counter-ions back in explicitly and the discrepancy resolves, and the way it resolves points straight at the controlling variable. Across twenty-seven microsolvated clusters the response sorts by the charge rule. The reason is simple electrostatics: a complex's incremental fluoride affinity falls monotonically as its own charge grows more negative, crossing zero near $1-$, and past that point the counter-charge a finite cluster can supply no longer overcomes anion-anion repulsion within the first shell. Sorted this way the three actinides behave alike, and each appears on both sides of the boundary only to the extent that its accessible coordination numbers put it there.

The practical consequence is that any predictive treatment of fuel-salt speciation has to include the alkali environment explicitly. An isolated actinide-fluoride complex, however accurately computed, will misassign the dominant species and will not transfer between carrier salts of differing alkali composition. It also marks a limit on the cluster approach itself, and our own results show it. The U(III) series sits entirely on the holding side of the boundary, and none of the four clusters climbs at either shell. Experiment puts U$^{3+}$ near CN~$\approx6$ in fluoride melts, and the finite cluster recovers neither the size nor the direction of that shift. We capture the mechanism and the ordering but underestimate the screening, and we underestimate it worst exactly where the complex is most negative. Reproducing U(III) speciation in the melt will take either a much larger explicit environment or a periodic description.

Thorium and uranium still differ substantially in electronic structure, and that difference governs everything except the solvation response. The $5f^0$ thorium center bonds essentially ionically and keeps a wide frontier gap that barely moves with coordination, so its speciation follows from electrostatics and entropy alone and is comparatively easy to model. Uranium behaves quite differently. Its bonding carries a genuine $5f$ covalent contribution, visible in the elevated Mayer bond orders at matched coordination, and its partially occupied $5f$ manifold sits pinned at the frontier. That covalency is what binds the U(IV) first shell most tightly of the three, and it shows up in the exchange barriers rather than in the solvation response. The frontier manifold is why the uranium electronic structure is coordination-dependent and, at the highest coordination, out of reach of a single-reference description. For reactor chemistry the lesson is that the thorium and uranium components of a Th/U fuel cannot be assumed to behave alike: uranium speciation is the harder of the two to predict and the more sensitive to local coordination.

The most consequential result is what this coordination dependence does to the redox thermodynamics. Across the bare series the intrinsic adiabatic reduction energy of the U(IV)/U(III) couple swings by roughly $11$~eV between CN~4 and CN~7, and it changes sign on the way: at low coordination U(III) is the stable state and the complex is reducing, at high coordination U(IV) is favored and the complex is oxidizing. The fluoride shell is not a small correction to a background potential. It is the dominant variable, and it tracks the frontier-gap collapse of Section~\ref{sec:physbasis} step for step, the U(III) gap falling from $2.65$~eV at CN~4 to complete closure at CN~7 over exactly the range in which the reduction energy turns positive. Since U(IV)/U(III) is the principal redox-active pair in a uranium-bearing fluoride melt, the redox state of the fuel, and with it the corrosion potential the salt presents to structural alloys, is not a fixed property of the salt at all. It shifts with the local coordination environment of the dissolved uranium. A reactor-scale model that treats $E^\circ$ as a composition-independent constant is working with the wrong kind of quantity.

The fluoride-exchange barriers are low throughout ($4$ to $17$~kcal\,mol$^{-1}$, with low-coordination additions barrierless), indicating that the coordination sphere can rearrange readily. U(III) exchanges fastest, Th(IV) slowest, and U(IV) lies in between, consistent with the corresponding frontier-gap trends. The reduced, small-gap uranium center is electronically softer and more willing to rearrange its shell, whereas closed-shell thorium is more rigid. The position of the $5f$ manifold relative to the frontier therefore influences structure, redox energetics, and exchange kinetics alike. The low barriers also allow solvated clusters to change coordination during ordinary geometry relaxation without an imposed reaction coordinate; every cluster that gained or shed fluoride did so in this way. This lability indicates that the observed coordination changes reflect relative solvation preferences rather than kinetic trapping.

That combination created an obstacle to solvated redox energetics we had not anticipated. A reduction energy compares two oxidation states at the same coordination number, but the two members of a redox pair necessarily differ by one unit of charge, and charge is exactly what sets the solvation response. The pair is charge-mismatched by construction and sits at adjacent points on the response curve. The $[\text{UF}_5]^-$/$[\text{UF}_5]^{2-}$ pair of Section~\ref{sec:redox} shows it happening: prepared at matched coordination, it does not stay matched through relaxation. Our bare complexes can hold integer coordination because they have no shell to exchange with. The solvated ones cannot. Any cluster-based treatment of solvated actinide redox will run into this, and constraining both oxidation states is not an obvious fix, since it forces a match at a coordination number neither species adopts.

The single-reference description also fails at high coordination, and while we can characterize that failure well, we cannot resolve it. Three independent diagnostics agree. The uranium spin population of the solvated U(III) clusters drops from $3.10$ at CN~4 and $3.08$ at CN~5 to $2.08$ at CN~6 and CN~7 once the second shell is present, a threshold rather than a gradual decline. The uranium Mulliken charge rises across the same threshold, from between $+1.76$ and $+1.95$ in the clusters that keep the formal occupation to $+2.43$ and $+2.28$ in those that do not, so the electron missing from the spin density reappears as charge transferred off the metal. And an atomic-valence projection onto the uranium $5f$ shell at CN~7 finds six of the seven $5f$-derived orbitals lying above the Fermi level of the reference determinant, which is the same depopulation expressed in the orbital space and obtained without any correlated calculation.

Whether that ligand-reduced state is physical or an artifact of the single determinant, we cannot settle here. An active space selected by atomic-valence character gives a clean seven-orbital $5f$ set at CN~5 and CN~6, but at CN~7 the same construction fails at the standard threshold precisely because the $5f$ manifold has been emptied, and the space we can obtain at a relaxed threshold admits two predominantly ligand orbitals. Worse, orbital optimization in these spaces is unstable: across eight calculations at three coordination numbers and every damping setting tested, the optimization drives the active-space occupations to integers and the space collapses to a single determinant, so the fractional occupations and multiconfigurational expansions we get at fixed orbitals do not survive relaxing those orbitals. The space as constructed is not stable, and it cannot support any statement about how strong the correlation in these states is. What the analysis does give us is the shape of the problem. The $5f$ manifold in these clusters is hybridized into the ligand field rather than sitting cleanly above it, which is at once why the single determinant collapses, why a valence $5f$ active space cannot be defined at CN~7, and why one defined by lowering the threshold comes out contaminated. What these clusters need is an active space spanning the ligand orbitals the $5f$ manifold mixes with, orbital-optimized and carrying dynamic correlation.

For anyone modeling reduced actinides in fluoride melts, the practical reading does not depend on that uncertainty, and it may be the most transferable result here. Single-reference DFT is well behaved for U(III) at low coordination and fails in the range the melt actually occupies. The calculations converge cleanly, with tight gradients and textbook convergence behavior, to a state whose spin population and metal charge are inconsistent with the formal occupation. The diagnostics that catch it are cheap, and they should be run as a matter of course.

Several limitations bound this study. The thermochemistry applies a gas-phase rigid-rotor/harmonic-oscillator treatment to species that live in a condensed phase, which puts a systematic error into the absolute free energies. The comparisons that carry our conclusions are between coordination numbers at fixed oxidation state, where that error largely cancels, but the absolute free energies should not be read as melt values.

The reduction energies are intrinsic quantities, computed without an absolute-electrode reference and without the differential solvation of the two redox partners. They are not electrode potentials, and they are not directly comparable to the measured values of Straka \textit{et al.} \cite{straka2010electrochemistry}, Jiang \textit{et al.} \cite{jiang2017electrochemical}, or Clayton \textit{et al.} \cite{clayton1974electrochemical}; what this work establishes is the coordination dependence, not the absolute scale. The second omission bears on the result directly, since the two members of the couple differ by one unit of charge and are therefore solvated to different extents, with the difference growing as the shell expands. A Born estimate over the dielectric constants appropriate to a fluoride melt ($\varepsilon = 1.9$-$3.0$) puts the differential solvation term negative at every coordination number and growing in magnitude with it, from about $-0.9$ to $-1.3$~eV at CN~4 to between $-6.5$ and $-9.2$~eV at CN~7. It opposes the computed trend rather than generating it, so solvation cannot be the origin of the coordination dependence: it acts in the wrong direction. It is not a small correction, though. Applied to the intrinsic energies it cuts the CN~4 to CN~7 swing from $11.1$~eV to $5.5$~eV at $\varepsilon = 1.9$ and to $3.2$~eV at $\varepsilon = 3.0$. The reduction energy stays monotonic in coordination number and still changes sign in both cases, but the sign change moves from between CN~4 and CN~5 to between CN~5 and CN~6 at the higher dielectric constant. The swing quoted here should therefore be read as an upper bound on the coordination dependence rather than as its melt value, and the bare-ion figures are inflated by the same finite-basis binding of the highly charged anions noted in Section~\ref{sec:Compdetails}, which a continuum partially removes. The qualitative conclusion, that coordination number is a redox control variable of first importance in these melts, is unaffected; its magnitude is bounded rather than determined by what we have done. A cavity-based treatment of the same correction is sensitive to the assumed dielectric constant and, at CN~6 and CN~7, to the state collapse of Section~\ref{sec:redox}, so we quote the Born estimate as a bound and not as a correction to apply.

Inner-sphere reorganization energies were attempted from a four-point construction and are not reported as a result: three of the four values come out negative, which a reorganization energy cannot be, and the construction has clearly not converged to a meaningful decomposition for anions of this charge.

Finally, the scalar-relativistic ZORA treatment omits spin-orbit coupling, which splits the $5f$ manifold of the open-shell uranium centers. The free-ion $5f$ splitting is at most of order $1$~eV, and a ligand field quenches it substantially, the more so the higher the angular momentum of the shell \cite{wei2025valence}. The coordination and bonding trends here rest on scalar energetics, orbital populations, and gaps of order $1$~eV and above, and a residual splitting of that size does not affect them. The smallest gaps are another matter. The U(III) frontier gap at CN~7 has closed entirely within the resolution of the scalar calculation, so it should be read as an indicator of near-degeneracy rather than as a quantitative excitation energy, and any quantitative redox energetics built on the high-coordination gaps would need a spin-orbit treatment. Note too that the all-electron SARC-ZORA basis carries the U~$6p$ semi-core shell explicitly instead of folding it into a pseudopotential, so the destabilization of the fluoride valence levels by that shell is present in our scalar description. What a scalar treatment cannot reproduce is the spin-orbit splitting the same core-valence mixing induces in the ligand-derived levels, which is an order of magnitude larger at $6p$ than at $5f$ \cite{wei2025valence}.

\section{Conclusion}
\label{sec:conclusion}

We have used relativistic density functional theory with explicit LiF microsolvation to characterize the coordination chemistry, redox energetics, and fluoride-exchange kinetics of Th(IV), U(IV), and U(III) in fluoride carrier salts, with particular attention to U(III), which has remained largely uncharacterized in fluoride melts. Five results stand out.

First, what sets U(III) apart from the higher oxidation states is its added $5f$ electron. It expands the coordination sphere and reverses the actinide contraction seen between Th(IV) and U(IV), and because it occupies a frontier orbital, the HOMO-LUMO gap collapses as coordination rises, from $2.65$~eV at CN~4 to complete closure at CN~7. That makes U(III) the most redox-active of the three. Second, the bare ions prefer low coordination (CN~5 for Th(IV) and U(IV), CN~4 for U(III)), well below experiment. Explicit LiF solvation narrows the gap through electrostatic screening, and it does so selectively: the sign of the response is fixed by the formal charge of the complex, so how a species behaves on solvation follows from that charge alone, with no reference to which actinide sits at the center. The anionic U(III) series lies entirely on the holding side of that boundary, which is why a finite cluster does not reproduce the experimental U(III) coordination. Third, coordination number then sets the redox thermodynamics. The intrinsic adiabatic reduction energy of the U(IV)/U(III) couple swings by roughly $11$~eV from CN~4 to CN~7 and changes sign: low coordination stabilizes U(III), high coordination stabilizes U(IV). A continuum estimate of the differential solvation of the two partners opposes that trend rather than producing it, and halves its magnitude without removing the sign change, so the intrinsic swing bounds the coordination dependence from above. Fourth, fluoride exchange is easy, with barriers of $4$ to $17$~kcal\,mol$^{-1}$, lowest for U(III) and highest for Th(IV), in the same order the frontier gaps predict. And because the two members of a redox couple differ by one unit of charge, and charge governs the solvation response, a solvated redox pair prepared at matched coordination does not stay matched under relaxation. That is a serious obstacle to cluster-based treatments of solvated actinide redox. Fifth, single-reference DFT loses the formal $5f^3$ configuration of the solvated U(III) clusters above a sharp threshold between CN~5 and CN~6, and three independent measures corroborate the loss: the metal spin population, the metal charge, and the position of the $5f$ manifold relative to the Fermi level. Whether the resulting ligand-reduced state is physical or an artifact of the single determinant, we could not establish. Active spaces selected by atomic-valence character are not stable under orbital optimization in these clusters, collapsing to a single determinant in every calculation we attempted.

These results establish the local coordination environment, set by the alkali counter-ions of the melt and by the charge of the complex rather than by the actinide itself, as the variable linking structure, electronic structure, redox thermodynamics, and exchange kinetics in the fluoride fuel salt. For reactor chemistry the consequence is that the redox potential of the fuel is not a constant to be measured once. It is a function of the local coordination environment, and any model that predicts it has to resolve that environment explicitly.

Two directions are open for exploration. The collapsed high-coordination solvated U(III) states need an orbital-optimized treatment carrying dynamic correlation, in an active space large enough to span the ligand orbitals the $5f$ manifold hybridizes with; the instability we ran into is itself evidence that a purely $5f$ valence space is the wrong choice for these clusters, and it locates the difficulty specifically enough to be addressed. Extending the analysis to chloride carrier salts, where the weaker ligand field should shift both the coordination preferences and the redox window, is the subject of ongoing study. Part of the value of this work is diagnostic: it marks where single-reference DFT succeeds for actinide-halide melts and where it breaks down, and it traces the breakdown to specific, addressable causes, the frontier degeneracy of the high-coordination $5f$ manifold and the lability of the solvated coordination sphere, rather than leaving it an unexplained failure.

\section{Acknowledgements}
We thank Prof.\ Brenda Rubenstein for her support and guidance throughout this work, and Brown University's Center for Computation and Visualization for access to the Oscar high-performance computing cluster and to the ORCA and Gaussian program packages.

\section{Data Availability}
All optimized structures, input files, harvested results and the analysis
scripts used to generate every table and figure in this work are available at
\url{https://doi.org/10.5281/zenodo.21849466}.

\bibliographystyle{unsrt}
\bibliography{refs12}

\end{document}